\documentclass[secnumarabic, graphics,floatfix, nofootinbib,tightenlines,nobibnotes, aps, prl, 12pt]{revtex4-1}
\usepackage{graphicx}
\usepackage[english]{babel}
\usepackage[utf8]{inputenc}
\usepackage{amsmath,amssymb}
\usepackage{amsfonts}
\usepackage{multirow}

\def\be{\begin{equation}}
\def\ee{\end{equation}}
\def\bea{\begin{eqnarray}}
\def\eea{\end{eqnarray}}

\def\ave#1{\langle#1\rangle}

\begin{document}

\title{On the boundary condition and related instability in the Smoothed Particle Hydrodynamics}

\author{Chong Ye$^{1}$}
\author{Philipe Mota$^{2}$}
\author{Jin Li$^{1*}$}
\author{Kai Lin$^{3,4}$}
\author{Wei-Liang Qian$^{4,5,6}$}

\affiliation{$^{1}$ College of Physics, Chongqing University, 401331, Chongqing, China}
\affiliation{$^{2}$ Centro Brasileiro de Pesquisas Físicas, 22290-180, Rio de Janeiro, RJ, Brazil}
\affiliation{$^{3}$ Institute of Geophysics and Geoinformatics, China University of Geosciences, 430074, Wuhan, Hubei, China}
\affiliation{$^{4}$ Escola de Engenharia de Lorena, Universidade de S\~ao Paulo, 12602-810, Lorena, SP, Brazil}
\affiliation{$^{5}$ Faculdade de Engenharia de Guaratinguet\'a, Universidade Estadual Paulista, 12516-410, Guaratinguet\'a, SP, Brazil}
\affiliation{$^{6}$ School of Physical Science and Technology, Yangzhou University, 225002, Yangzhou, Jiangsu, China}

\date{July 31, 2019}          

\begin{abstract}
In this work, we explore various relevant aspects of the Smoothed Particle Hydrodynamics regarding Burger's equation.
The stability, precision, and efficiency of the algorithm are investigated in terms of different implementations. 
In particular, we argue that the boundary condition plays an essential role in the stability of numerical implementation.
Besides, the issue is shown to be closely associated with the initial particle distribution and the interpolation scheme.
Among others, we introduce an interpolation scheme termed symmetrized finite particle method.
The main advantage of the scheme is that its implementation does not involve any derivative of the kernel function.
Concerning the equation of motion, the calculations are carried out using two distinct scenarios where the particles are chosen to be either stationary or dynamically evolved.
The obtained results are compared with those obtained by using the standard finite difference method for spatial derivatives.
Our numerical results indicate subtle differences between different schemes regarding the choice of boundary condition.
In particular, a novel type of instability is observed where the regular distribution is compromised as the particles start to traverse each other. 
Implications and further discussions of the present study are also addressed.

\end{abstract}

\maketitle                                

\section{I. Introduction}

The smoothed particle hydrodynamics (SPH)~\cite{sph-app-astro-01,sph-app-astro-02,sph-algorithm-review-01,sph-algorithm-review-02,sph-algorithm-review-03,sph-algorithm-review-04,sph-algorithm-review-05,sph-algorithm-review-06,sph-algorithm-review-10} is a meshfree, Lagrangian method for the partial different equations.
Distinct from traditional approaches such as the finite difference method (FDM) or finite element method, the SPH approximates a partial differential equation in terms of ordinary differential equations of freely moving interpolation points, known as SPH particles.
Subsequently, the infinite number of degrees of freedom of the original system is represented by a finite number of SPH particles, aiming to describe any arbitrary configuration of the continuum medium.
As one of the most important applications of the partial differential equation is to describe the dynamics of fluid, SPH method naturally finds extensive applications in the context of hydrodynamics, as inferred by the terminology.
Indeed, due to the significance of hydrodynamics, as well as recent advance in computing power, the SPH method is widely employed in many distinct areas of science and engineering.
Among traditional applications to fluid dynamics, the SPH method has been employed to investigate a great variety of topics.
In astrophysics, the SPH is implemented for the study of galaxy formation, supernova explosion~\cite{sph-algorithm-review-05,sph-algorithm-review-09,sph-algorithm-review-10,sph-app-astro-03}
In solid state mechanics~\cite{sph-app-01}, the method has been applied to problems such as metal forming~\cite{sph-app-02}, fracture and fragmentation~\cite{sph-app-03}, and shock wave propagation in solids~\cite{sph-app-04}.
Moreover, concerning the high energy nuclear physics, the hydrodynamic approach of heavy-ion nuclear collisions plays an essential role associated with RHIC and LHC~\cite{hydro-review-07,sph-review-1,hydro-review-06}.
The latter further inspires onging investigations of fluid/gravity duality~\cite{adscft-fluidgravity-01,adscft-fluidgravity-02,adscft-fluidgravity-04,adscft-fluidgravity-06}, one realization of the holographic principle~\cite{adscft-01,adscft-02}.
Numerical simulations based on implementations of SPH algorithm~\cite{sph-corr-1,hydro-vn-2,sph-corr-4,hydro-corr-2,sph-corr-ev-4} have shown to give reasonable description on various experimental observables.
The algorithm is also employed in chemistry where a variety of chemical evolutions simulations has been implemented in terms of the SPH method.

The SPH was initially suggested to investigate astrophysical problems by Gingold and Monaghan~\cite{sph-app-astro-02} and by Lucy~\cite{sph-app-astro-01}.
In the standard implementation, each SPH particle is assigned a finite size, $h$, over the extension of which the relevant physical properties are smoothed in terms of a kernel function. 
By kernel function, the contribution of each particle is weighted regarding their respective distance to the coordinate position of interest.
As a consequence, a physical quantity at a given position is evaluated by summing over all the contributions from all the relevant SPH particles in the neighborhood within reach of the kernel function.
In terms of the above interpolation regarding SPH particles, a partial differential equation is transformed into a system of ordinary differential equations of the SPH degrees of freedom.

The SPH algorithm is a meshfree approach, which implies several essential features.
First, the method is suited to handle problems with sophisticated boundary conditions, inclusively, scenarios with free boundary.
The algorithm can achieve exact advection mostly in a trivial manner, which is a rather difficult task for any Eulerian grid-based approaches. 
Meanwhile, the absence of the structured computational mesh, by and large, simplifies model implementation.
In particular, the spatial derivatives are implemented in a flexible and straightforward fashion, independent of any grid-based configuration.
Besides, one of the most remarkable features of the SPH method is that it naturally guarantees the conservation law. 
In other words, independent of any specific application, the algorithm rigorously conserves quantities such as mass, energy, momentum, and angular momentum by construction.
Also, a certain degree of re-meshing takes place automatically as time evolves, as a consequence of the kernel function.
It is distinct from other Lagrangian grid-based methods where the grid can sometimes become arbitrarily disordered in time~\cite{sph-algorithm-review-07}.
Moreover, when appropriately configured, it is understood that the computational cost of numerical simulations per SPH particle is usually less than that per cell of grid-based simulations. 
This is because the resolution of the method follows naturally the mass flow rather than the volume as in the case of most grid-based techniques.
It happens as each SPH particle carries a fixed amount of mass and dislocate itself according to the flow velocity, and subsequently, the density only goes up related to an increase in the local matter concentration~\cite{sph-algorithm-review-07}.
To be more specific, it is the case especially when the metric of interest is associated with the relevant density, and thus an automatical refinement on density is achieved.
Furthermore, the SPH method is highly flexible and can be combined with other algorithms straight forwardly~\cite{sph-algorithm-review-09}. 

However, the SPH formulation also faces a few challenges.
Several notable problems potentially hamper further exploiation of the method, such as accuracy, tensile instability~\cite{sph-algorithm-tensile-insti-01,sph-algorithm-tensile-insti-02,sph-algorithm-CSPM-03}, pairing instability~\cite{sph-algorithm-pairing-inst-01}, the kernel as well as particle consistency~\cite{sph-algorithm-review-06,sph-algorithm-10,sph-algorithm-CSPM-01,sph-algorithm-FPM-01,sph-algorithm-09}, and artificial viscosity~\cite{sph-algorithm-art-viscosity-01}.
From a physical viewpoint, tensile instability is triggered when SPH particles become mutually attractive due to, by and large, the presence of negative pressure. 
The eventual instability manifested regarding particle distribution, known as tensile instability, can be viewed as equivalent to the mesh distortion in a mesh-based scenario, is typically found in simulations of elastic material under tension.
One possible approach to deal with such instability associated with irregular particle distribution is to sacrifice exact conservation of the algorithm by a slim amount~\cite{sph-algorithm-tensile-insti-03}.
Besides, other approaches were proposed to handle the instability by improving the estimation of spatial gradients, such as the corrected derivatives method~\cite{sph-algorithm-tensile-insti-04}.
As the number of neighbors of an individual SPH particle increases, the ratio of the particle spacing to the smoothing length decreases.
However, for standard ``Bell-shaped" or ``Gaussian-like" kernels, their first derivative, and equivalently the force between SPH particles, vanishes at the origin. 
Consequently, particle pairing occurs.
This problem is relatively benign, as it usually does not lead to any severe consequence.
It can mostly be avoided by just using an adequate initial particle distribution with a smaller number of neighbors or adopting a short-range kernel function.
Concerning kernel and particle consistency, much effort has been devoted to restore the consistency and subsequently improve accuracy.
Monaghan derived a symmetrization formulation for the spatial gradient term~\cite{sph-algorithm-review-01} consistent with the variational approach, which is shown to provide better accuracy.
Randles and Libersky proposed a normalization scheme for the density approximation~\cite{sph-algorithm-10}.
The concept is further developed by Chen et al.~\cite{sph-algorithm-CSPM-01,sph-algorithm-CSPM-02,sph-algorithm-CSPM-03} in terms of a corrective smoothed particle method (CSPM) as well as by Liu et al.~\cite{sph-algorithm-FPM-01} in terms of a finite particle method (FPM). 
For some physical phenomena such as shockwave, the entropy is increased as a physical consequence.
In fact, unlike in a Eulerian scheme where numerical dissipation is an intrinsic feature related to the discretization process and mostly unavoidable, diffusion terms in SPH must be inserted explicitly, and as a result, their physical content is transparent.
In this regard, it has been argued that it is essential to supply an adequate amount of artificial viscosity which works for not only scenarios where artificial viscosity is required but also for other cases where its presence ought to be minimized~\cite{sph-algorithm-art-viscosity-02,sph-algorithm-art-viscosity-03}.

As a matter of fact, there does not exist a single numerical method which performs satisfactorily well at all different aspects.
Therefore, in practice, the choice of the best numerical approach can often depend entirely on which feature of the problem is more relevant to the present study.
Nonetheless, stability is usually the primary concern of a numerical scheme, while precision and efficiency are two essential performance metrics in most cases.
Our present study thus aims to provide a quantitative analysis of these aspects of the SPH method.
In particular, we explore the role of boundary condition, as well as its relationship with initial particle distribution and interpolation scheme.
Regarding the specific problem we choose to investigate, the equation of motion is implemented using two distinct scenarios, where the SPH particles are selected to be either stationary or allowed to evolve dynamically in time.
The first scenario is shown to be an excellent indicator of the efficiency and accuracy of the individual interpolation algorithm.
The second scenario, on the other hand, resides on a more general basis and is subsequently found to be sensitively associated with the stability of different implementations.

The paper is organized as follows.
In the following section, we briefly discussed the main features of different implementations of the SPH algorithms involves in the present study, namely, the standard SPH, CSPM, FPM, and the symmetrized finite particle method (SFPM).
The calculations are carried out in section III and IV where Burger's equation is chosen to be the object of the numerical simulations.
The results are then compared against the analytic solutions.
Two specific scenarios for the equation of motion are made use of with either stationary and dynamical SPH particles.
In section III we show the results concerning the first scenario, where we mainly focus on the precision and efficiency of different interpolation schemes.
The second scenario is studied in Section IV, which is mostly related to the stability of the algorithm regarding the boundary conditions.
We discuss how the boundary condition affects the stability of the numerical calculations for different schemes.
Also, a type of novel instability is observed where SPH particles are found to traverse each other, which, in turn, undermine the regular particle distribution.
The last section is devoted to further discussions and concluding remarks.

\section{II. The SPH method} \label{Sect2}

In the present study, we mainly focus on four different SPH interpolation schemes, namely, standard SPH, CSPM, FPM, and SFPM.
In what follows, what briefly review these schemes concerning their derivations. 
For the spatial distribution of a given physical quantity, $f(x)$, all the above schemes can be viewed as specific implementations to interpolate the distribution by using a finite number of SPH particles as well as the kernel function $W$.
The latter is assumed to be an even function and its normalization reads
\begin{eqnarray}
\int dx W(x-x_a) = 1 .
\label{eqWnorm}
\end{eqnarray}
The Taylor series expansion of $f(x)$ reads
\begin{eqnarray}
f(x) = \sum_{n=0} \frac{ (x-x_a)^n_i }{n!}(\partial^n_i f)_{x_a} .
\label{eq1}
\end{eqnarray}
We proceed by multiplying both sides by the kernel function $W(x-x_a)$ and integrating over $x$, we have
\begin{eqnarray}
\int dx f(x) W(x-x_a) = \sum_{n=0} \frac{ (\partial^n_i f)_{x_a} }{n!} \int dx (x-x_a)^n_i W(x-x_a) .
\end{eqnarray}
As one keeps the first term on the r.h.s. of the above equation and discard the higher order ones, one obtains
\begin{eqnarray}
\int dx f(x) W(x-x_a) = f_a \int dx W(x-x_a) .
\label{eq2}
\end{eqnarray}

By substituting the normalization condition Eq.(\ref{eqWnorm}), we obtain the standard SPH formula, namely,
\begin{eqnarray}
f_a = \int dx f(x) W(x-x_a) = \sum_b \frac{\nu_b f_b}{\rho_b} W(x_b-x_a) . \label{stdSPH}
\end{eqnarray}
We note that the particle approximation is utilized, where a summation enumerating all SPH particle is used to approximate the integral.
Here $\nu_b$ is the ``mass" of the $b$-th particle, usually it is a given value assign to the particle at the beginning of the algorithm.
$\rho_b$ is the density of the SPH particles, if $f$ itself is the density, one has
\begin{eqnarray}
\rho_a = \sum_b {\nu_b} W(x_b-x_a) . \label{Eqrhoa} 
\end{eqnarray}
In practice, the ratio $\frac{f_b}{\rho_b}$ can be determined by the physical properties of the system in question.
For instance, if $f$ represents pressure and $\rho$ is the number density, their ratio might be determined by the equation of state~\cite{sph-review-1}.
Otherwise, $f_b$ and $\rho_b$ can be that evaluated for the $b$-th particle at the last time step.
The first and second order derivatives can be obtained by carrying out spatial derivative of Eq.(\ref{stdSPH}), and are found to be
\begin{eqnarray}
f_{a,x} &=& \frac{1}{\rho_a}\sum_b {\nu_b (f_a-f_b)} W'(x_b-x_a) ,\nonumber \\
f_{a,xx} &=& \frac{1}{\rho_a}\sum_b {\nu_b (f_{a,x}-f_{b,x})} W'(x_b-x_a) ,\label{stdSPH2}
\end{eqnarray}
where symmetric formulism~\cite{sph-algorithm-review-01} has been employed.

One might be aware of the fact that in the particle approximation of Eq.(\ref{eqWnorm}), the kernel function is not precisely identical to the Dirac $\delta$-function, and instead employ the following modified particle approximation in the place of Eq.(\ref{eq2}),
\begin{eqnarray}
\int dx f(x) W(x-x_a) \rightarrow \sum_b \frac{\nu_b f_b}{\rho_b} W(x_b-x_a) \\
\int dx W(x-x_a) \rightarrow \sum_b \frac{\nu_b }{\rho_b} W(x_b-x_a) .
\end{eqnarray}
One finds
\begin{eqnarray}
f_a = \frac{\sum_b \frac{\nu_b f_b}{\rho_b} W(x_b-x_a)}{\sum_b \frac{\nu_b}{\rho_b} W(x_b-x_a)} . \label{EqCSPM}
\end{eqnarray}
Eq.(\ref{EqCSPM}) is known as CSPM in literature, which can be shown to preserve the zeroth order kernel and particle consistency~\cite{sph-algorithm-review-06}.
The first order and second order derivatives can be derived by retaining until the second order term on the r.h.s. of Eq.(\ref{eq1}), multiplying both sides by $W'$ and $W''$, and making use of the symmetry of these functions.
One finds
\begin{eqnarray}
f_{a,x} &=& \frac{\sum_b \frac{\nu_b (f_b-f_a)}{\rho_b} W'(x_b-x_a)}{\sum_b \frac{\nu_b (x_b-x_a)}{\rho_b} W'(x_b-x_a)} , \nonumber\\ \label{EqCSPM2}
f_{a,xx} &=& \frac{\sum_b \frac{\nu_b (f_b-f_a)}{\rho_b} W''(x_b-x_a)-f'_a\sum_b \frac{\nu_b (x_b-x_a)}{\rho_b} W''(x_b-x_a)}{\frac12 \sum_b \frac{\nu_b (x_b-x_a)^2}{\rho_b} W'(x_b-x_a)} .
\end{eqnarray}

In fact, one may proceed further by generalizing the above arguments to higher order in a slightly more consistent fashion. 
This is achieved by consistently expanding the r.h.s. of Eq.(\ref{eq1}) to the second order, namely,
\begin{eqnarray}
\int dx f(x) W(x-x_a) = f_a \int dx W(x-x_a) + f_{a,x} \int dx (x-x_a) W(x-x_a) \nonumber\\
+\frac12 f_{a,xx} \int dx (x-x_a)^2 W(x-x_a) . \label{philipe1}
\end{eqnarray}
Similarly, one may again replaces $W(x-x_a)$ by $W'(x-x_a)$ and $W'(x-x_a)$, and subsequently obtains
\begin{eqnarray}
\int dx f(x) W'(x-x_a) = f_a \int dx W'(x-x_a) + f_{a,x} \int dx (x-x_a) W'(x-x_a) \nonumber\\
+\frac12 f_{a,xx} \int dx (x-x_a)^2 W'(x-x_a) , \nonumber\\
\int dx f(x) W''(x-x_a) = f_a \int dx W''(x-x_a) + f_{a,x} \int dx (x-x_a) W''(x-x_a) \nonumber\\
+\frac12 f_{a,xx} \int dx (x-x_a)^2 W''(x-x_a)  . \label{philipe2}
\end{eqnarray}
The present case is different from that of CSPM scheme, since now Eqs.(\ref{philipe1}-\ref{philipe2}) furnishes a system of coupled equations, as follows
\begin{eqnarray}
\left[\begin{matrix}
\ave{f}_a \\
\ave{f}_{a,x} \\
\ave{f}_{a,xx}
\end{matrix}\right]
= 
\left[\begin{matrix}
 \ave{1}_a & \ave{\Delta x}_a & \ave{\Delta x^2}_a \\
 \ave{1}_{a,x} & \ave{\Delta x}_{a,x} & \ave{\Delta x^2}_{a,x} \\
 \ave{1}_{a,xx} & \ave{\Delta x}_{a,xx} & \ave{\Delta x^2}_{a,xx}
\end{matrix}\right]
\left[\begin{matrix}
f_a \\
f_{a,x} \\
f_{a,xx}
\end{matrix}\right] , \label{philipe}
\end{eqnarray}
where the averages $\ave{\cdots}$ are defined as follows 
\begin{eqnarray}
\ave{f}_a &\equiv& \sum_b \frac{\nu_b f_b}{\rho_b} W_{ab} ,\nonumber\\
\ave{f}_{a,x} &\equiv& \sum_b \frac{\nu_b f_b}{\rho_b} W'_{ab} ,\nonumber\\
\ave{f}_{a,xx} &\equiv& \sum_b \frac{\nu_b f_b}{\rho_b} W''_{ab} . \label{FPM}
\end{eqnarray}
By inverting the above matrix equation, one obtains for the desired expressions for the FPM scheme, which read
\begin{eqnarray}
f_{a}&=&
\frac{1}{A}\Big(\ave{f}_{a} \big(\ave{\Delta x}_{a,x}\ave{\Delta x^{2}}_{a,xx}-\ave{\Delta x^{2}}_{a,x}\ave{\Delta x}_{a,xx}\big)\nonumber\\
       &+&\ave{f}_{a,x}\big(\ave{\Delta x^{2}}_{a}\ave{\Delta x}_{a,xx}-\ave{\Delta x}_{a}\ave{\Delta x^{2}}_{a,xx}\big)\nonumber\\
       &+&\ave{f}_{a,xx}\big(\ave{\Delta x}_{a}\ave{\Delta x^{2}}_{a,x}-\ave{\Delta x^{2}}_{a}\ave{\Delta x}_{a,x}\big)\Big) , \nonumber\\
f_{a,x}&=&\frac{1}{A}\Big(\ave{f}_{a}\big(\ave{\Delta x^{2}}_{a,x}\ave{1}_{a,xx}-\ave{1}_{a,x}\ave{\Delta x^{2}}_{a,xx}\big)\nonumber\\
          &+&\ave{f}_{a,x}\big(\ave{1}_{a}\ave{\Delta x^{2}}_{a,xx}-\ave{\Delta x^{2}}_{a}\ave{1}_{a,xx}\big)\nonumber\\
          &+&\ave{f}_{a,xx}\big(\ave{\Delta x^{2}}_{a}\ave{1}_{a,x}-\ave{1}_{a}\ave{\Delta x^{2}}_{a,x}\big)\Big) , \nonumber\\
f_{a,xx}&=&\frac{1}{A}\Big(\ave{f}_{a}\big(\ave{1}_{a,x}\ave{\Delta x}_{a,xx}-\ave{\Delta x}_{a,x}\ave{1}_{a,xx}\big)\nonumber\\
            &+&\ave{f}_{a,x}\big(\ave{\Delta x}_{a}\ave{1}_{a,xx}-\ave{1}_{a}\ave{\Delta x}_{a,xx}\big)\nonumber\\
            &+&\ave{f}_{a,xx}\big(\ave{1}_{a}\ave{\Delta x}_{a,x}-\ave{\Delta x}_{a}\ave{1}_{a,x}\big)\Big) , \label{defDel}
\end{eqnarray}
with
\begin{eqnarray}
A&=&-\ave{\Delta x^{2}}_{a}\ave{\Delta x}_{a,x}\ave{1}_{a,xx}+\ave{\Delta x}_{a}\ave{\Delta x^{2}}_{a,x}\ave{1}_{a,xx}\nonumber\\
   &+&\ave{\Delta x^{2}}_{a}\ave{1}_{a,x}\ave{\Delta x}_{a,xx}-\ave{1}_{a}\ave{\Delta x^{2}}_{a,x}\ave{\Delta x}_{a,xx}\nonumber\\
   &-&\ave{\Delta x}_{a}\ave{1}_{a,x}\ave{\Delta x^{2}}_{a,xx}+\ave{1}_{a}\ave{\Delta x}_{a,x}\ave{\Delta x^{2}}_{a,xx} .\nonumber
\end{eqnarray}
The FPM scheme is understood to possess the zeroth and first-order particle consistency~\cite{sph-algorithm-review-06}.

It is worth mentioning that in deriving Eq.(\ref{philipe2}), one has the freedom to utilize any basis function.
Another particular choice, which is meaningful by its own right, is to make use of $(x-x_a)W(x-x_a)$ and $(x-x_a)^2W(x-x_a)$ in the place of $W'(x-x_a)$ and $W'(x-x_a)$.
The only difference is that the definitions in Eq.(\ref{FPM}) are replaced by
\begin{eqnarray}
\ave{f}_a &\equiv& \sum_b \frac{\nu_b f_b}{\rho_b} W_{ab} ,\nonumber\\
\ave{f}_{a,x} &\equiv& \sum_b \frac{\nu_b f_b(x_b-x_a)}{\rho_b} W_{ab} ,\nonumber\\
\ave{f}_{a,xx} &\equiv& \sum_b \frac{\nu_b f_b(x_b-x_a)^2}{\rho_b} W_{ab} . \label{sFPM}
\end{eqnarray}
We note that the alternative basis functions maintain the symmetries (even or odd with respect to the origin) of the integral kernel.
Meanwhile, the formalism has avoided using the derivatives of the kernel function, which turns out to be quite advantageous as shown below.
This last scheme was initially proposed in Ref\cite{sph-algorithm-FPM-05} and will be referred to as the symmetrized finite particle method (SFPM) in the present work.

To sum up, Eqs.(\ref{stdSPH}-\ref{stdSPH2}), Eq.(\ref{EqCSPM}), Eqs.(\ref{FPM}-\ref{defDel}) and Eqs.(\ref{defDel}-\ref{sFPM}) give the specific formalism for standard SPH, CSPM, FPM and SFPM.
In the remainder of the paper, these differences will be put into practice in the following two sections.

\section{III. Burgers' equation with stationary SPH particles: on precision and efficiency of different schemes} \label{Sect3}

In this section, we focus ourselves on the static aspect of the algorithm regarding the precision as well as the efficiency of different schemes.
By and large, the precision of a given scheme can be measured in terms of kernel consistency and particle consistency.
The former is determined by the $n$-th moment of the kernel functions, namely,
\begin{eqnarray}
C^n&:& \int (x-x')^n W(x-x')dx=0 ,\label{kCon}
\end{eqnarray}
for $n=0, 1, 2\cdots$.
The above conditions are rigorous when the interpolation is perfect.
When it is satisfied for a given order $n$, the corresponding SPH approximation is said to possess $n$-th order or $C^n$ consistency.

The particle consistency, on the other hand, is related to the fact that in practice, the integral in Eq.(\ref{kCon}) is implemented by a summation.
As a result, one requires
\begin{eqnarray}
C^n&:& \sum_b \frac{\nu_b(x_b-x_a)^n}{\rho_b}W(x_b-x_a)=0 .\label{pCon}
\end{eqnarray}

As mentioned above, the present section is primarily dedicated to investigating to what degree the SPH algorithm is capable of reproducing a well-defined distribution of certain physical quantity in terms of the SPH particles.
In this context, the precision and efficiency are affected by various factors, such as the specific form of the kernel function, the number of particles, and the interpolation scheme.
In what follows, we carry out numerical simulations while tuning parameters regarding these aspects.
We note that, for the time being, we will temporarily discard another vital factor, the stability of the algorithm.
In other words, we only concentrate on the precision and efficiency of specific problems where the stability is guaranteed or a minor issue. 
In this regard, we study the following one-dimensional Burgers' equation:
\begin{eqnarray}
\frac{\partial u}{\partial t}=\gamma \frac{\partial ^2u}{\partial x^2}-u\frac{\partial u}{\partial x} ,\label{BurgerEq}
\end{eqnarray}
where $\gamma$ is the viscosity coefficient.
The reason for our choice of Burgers' equation is as follows.
First, the equation possesses an analytic solution, corresponding to smooth initial conditions where $u$ vanishes on the boundary. 
Secondly, usually, the physical interpretation of the equation is the same as the Navier-Stokes equation. 
In other words, the Burgers' equation determines the temporal evolution of, $u$, the velocity of the fluid.
Therefore, the physical system is governed by two equations for the fluid velocity $u(x,t)$ and matter density $\rho(x,t)$, respectively.
The latter is nothing but the continuity equation of the fluid, which takes the following form in one-dimensional space,
\begin{eqnarray}
\frac{\partial \rho}{\partial t}=-\frac{\partial (\rho u)}{\partial x} .\label{rhoEq}
\end{eqnarray}
However, it is interesting to point out that one may simply view $u$ as an intensive physical quantity, and solve Eq.(\ref{BurgerEq}) alone independent of Eq.(\ref{rhoEq}).
This is possible since, for the specific case of the Burgers' equation, the equation of $u$ does not depend on $\rho$.
Therefore, when implementing the SPH scheme, we choose to assign the SPH particles to fixed spatial coordinates. 
Here, $u_a$ of the $a$-th particle is a function of time, determined by the r.h.s. of Eq.(\ref{BurgerEq}), while the coordinates of the SPH particles remain unchanged during the simulation.
Lastly, the boundary condition of the present problem is that the distribution attains zero at both end points, namely,
\begin{eqnarray}
u(0)=u(1)=0 .\label{BoundaryCondition}
\end{eqnarray}
But since the SPH particle do not dislocate in time, for the present section, one does not need to explicitly implement Eq.(\ref{BoundaryCondition}) into the code.
As shown in the next section, the latter is closely related to the issue of stability.
For simplicity, throughout our calculations, we employ the fifth order spline kernel function~\cite{sph-algorithm-08}.
We also note that the present application involves fixed boundary condition, while on the other hand, SPH algorithm is most advantageous for problems with generic free boundary condition, the resultant precision and efficiency of SPH is not expected to be superior to the outcome of traditional FDM.
In what follows, the different schemes discussed in the previous section will be applied by using the above setup.

\begin{figure}[ht]
\begin{tabular}{cc}
\vspace{-30pt}
\begin{minipage}{225pt}
\centerline{\includegraphics[width=300pt]{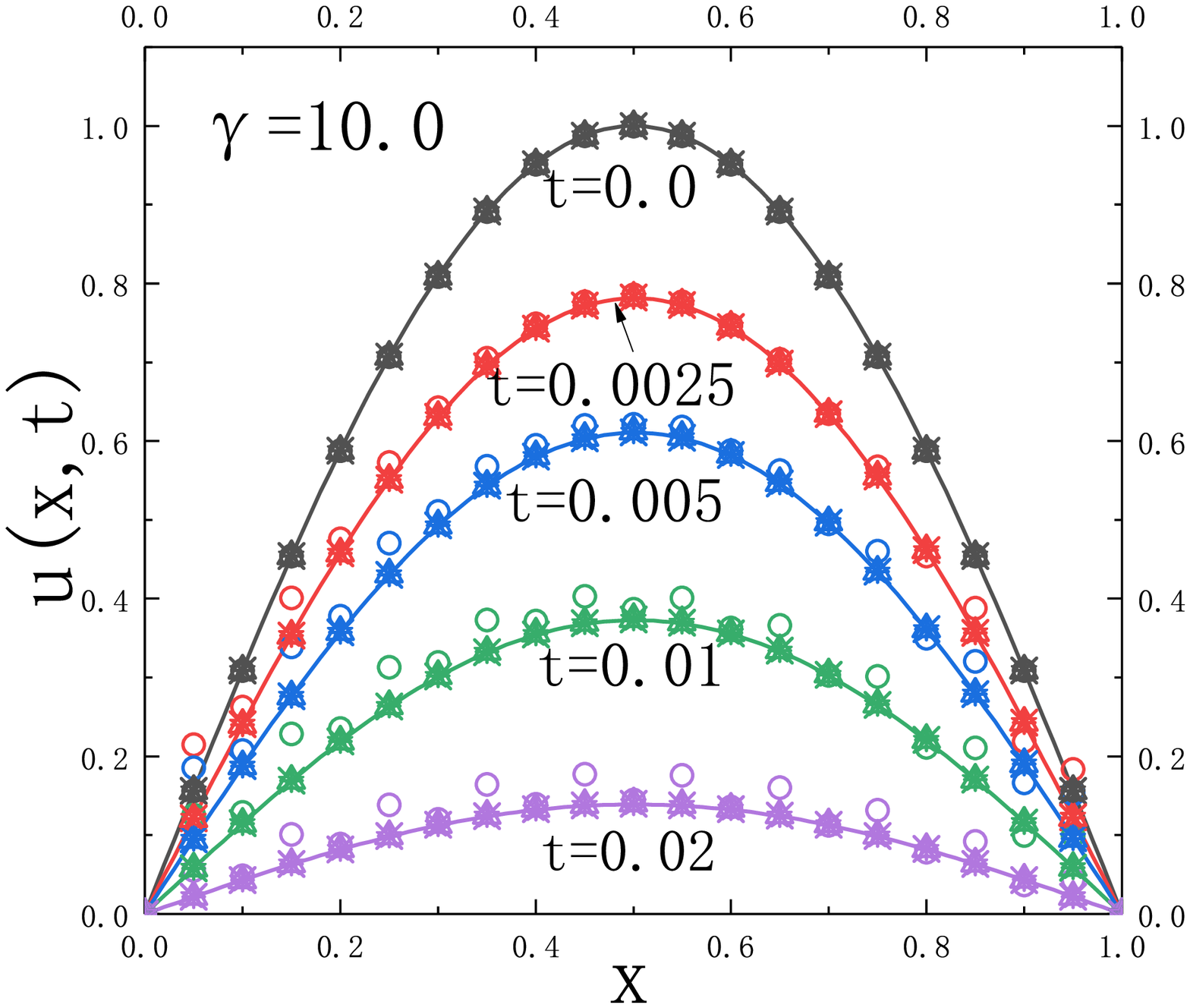}}
\end{minipage}
&
\begin{minipage}{225pt}
\centerline{\includegraphics[width=300pt]{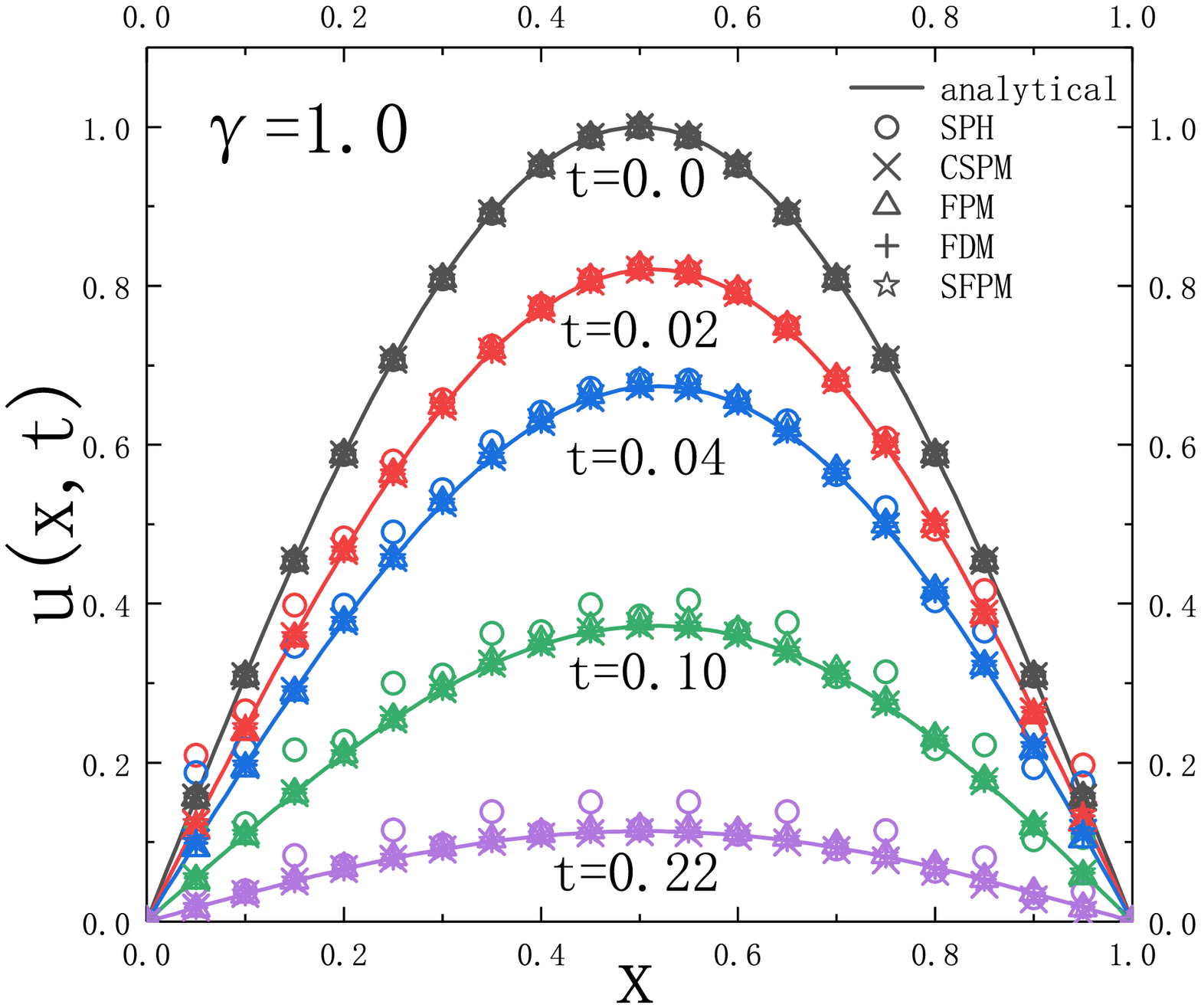}}
\end{minipage}
\\
\begin{minipage}{225pt}
\centerline{\includegraphics[width=300pt]{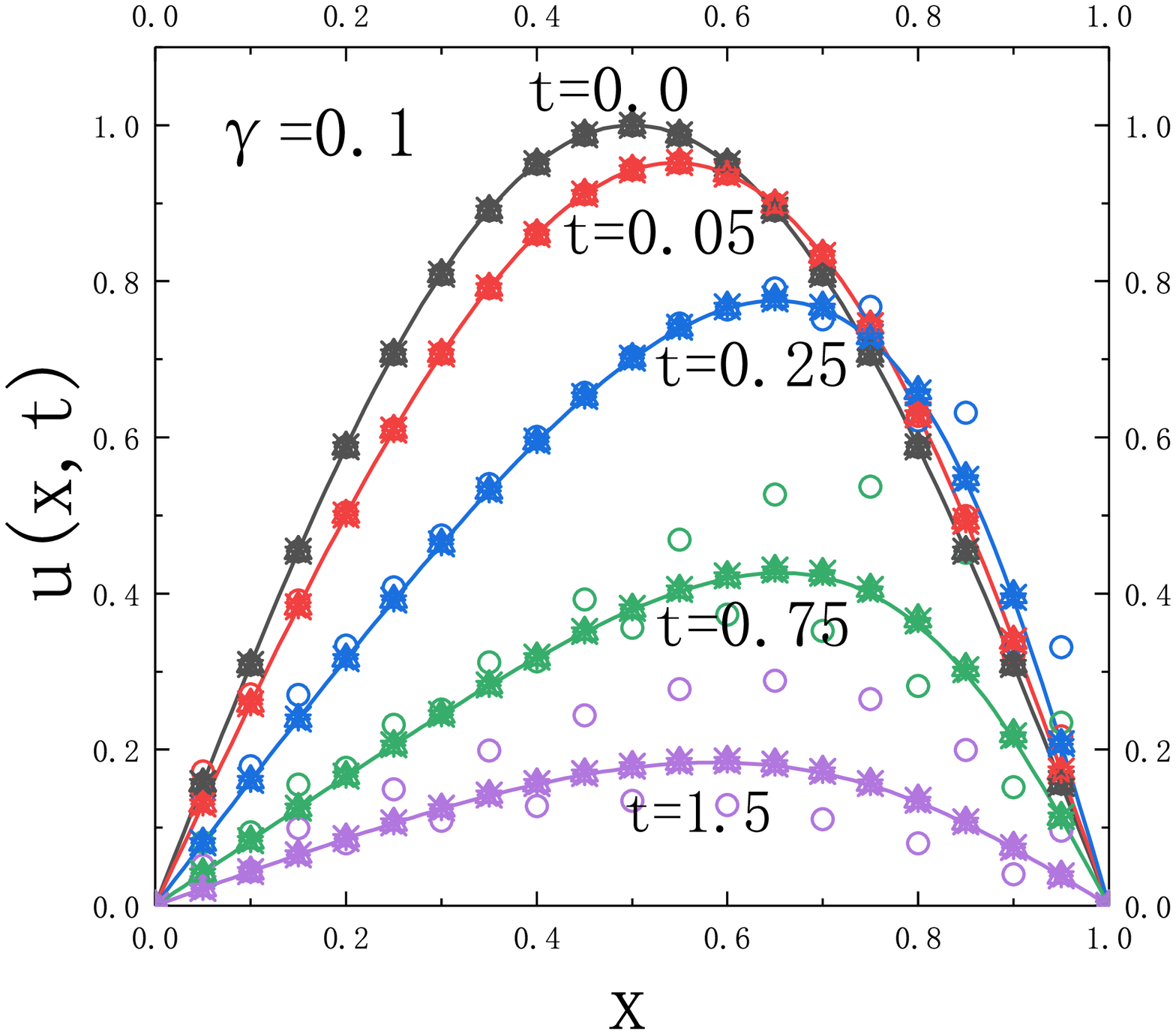}}
\end{minipage}
&
\begin{minipage}{225pt}
\centerline{\includegraphics[width=300pt]{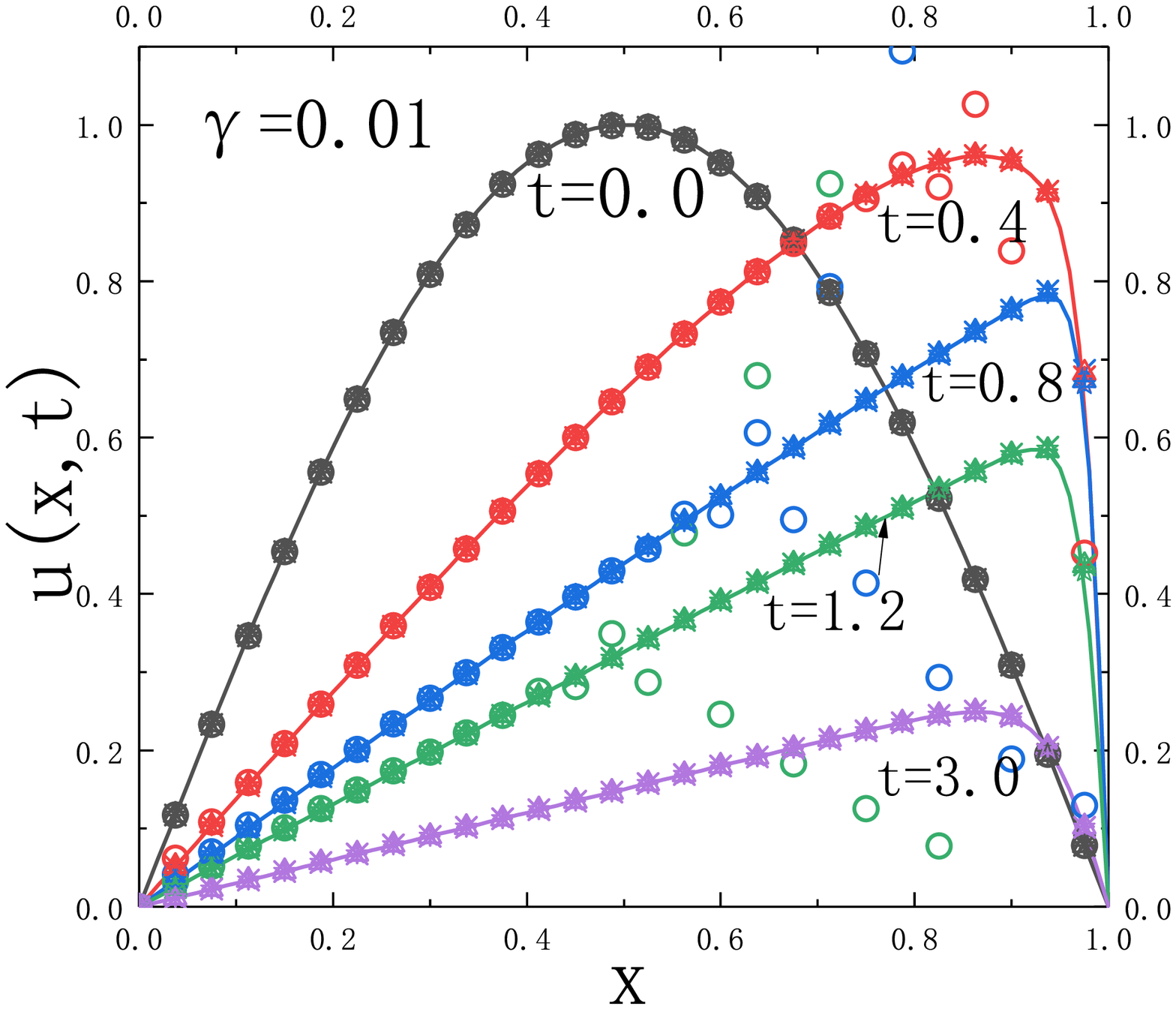}}
\end{minipage}
\end{tabular}
\renewcommand{\figurename}{Fig.}
\caption{(Color online) The numerical solutions of the Burgers' equation in comparison with the analytic ones. 
The calculations are carried out by using stationary SPH particles for standard SPH, CSPM, FPM, FDM, and SFPM, respectively.
The results are presented for different values of the viscosity coefficients $\gamma$.}
\label{temporalEvo-1}
\end{figure}
In Fig.\ref{temporalEvo-1}, we show the calculated temporal evolutions of the distribution in comparison with those obtained by using the FDM as well as the analytic solutions.
It is found that most methods give consistent and reasonable results for different viscosity coefficients.
Among all the interpolations schemes, the standard SPH shows the most substantial deviations from the analytic results.
To precisely identify the precision and efficiency of different schemes, we numerically evaluate the deviations from analytic results.
In Tab\ref{AreEffTable}, the averaged relative precision and efficiency for different numbers of SPH particles or grid numbers (for FDM) are summarized.
In our calculations, the particles are distributed uniformly, and the size of SPH particles, $h$, varies according to the total number of SPH particles.
To be specific, we choose $h=\Delta x$ where $\Delta x$ is the distance between neighbor particles.
In the calculations, the time interval is taken to be $\Delta t=h/500$ , and we consider $\gamma=1.0$.
The precision is estimated for the instant $t=1.0$ while regarding an average of the relative deviation
\begin{equation}\label{Are}
  P=\frac{\sum\limits_{i=1}^{N}(|f_{i}-f_{ana}|)/|f_{ana}|}{N} ,
\end{equation}
where $N$ is the total number of selected spatial points where the quality of the interpolation is evaluated by comparing to analytic result.
In our calculations, we have used seven evenly distributed spatial points for the evaluation.
As the efficiency, one additionally takes into account the CPU time consumption, $\Delta t$.
The efficiency of a given method is defined as follows
\begin{equation}\label{2}
  E=\frac{\Delta t}{\Delta t_{\mathrm SPH}} P ,
\end{equation}
which is measured with respect to that of the standard SPH method.

\begin{table}[!htbp]
  \centering
  \caption{The relative precision and efficiency of different methods.
The first row indicates the number of SPH particles or grids used in the calculations.}
  \label{AreEffTable}
  \begin{tabular}{|p{2.0cm}|p{1.9cm}|p{1.9cm}|p{1.9cm}|p{1.9cm}|p{1.9cm}|p{1.9cm}|p{1.9cm}|}
    \hline
    No. &  6  &11  &21  &51  &101  &161  &201 \\
    \hline
    $P_{\mathrm SPH}$&   $3.71\times10^{-1}$        &$1.63\times10^{-1}$       &$8.2\times10^{-2}$        &$3.5\times10^{-2}$        &$1.8\times10^{-2}$        &$1.1\times10^{-2}$   &$8.9\times10^{-3}$\\
    \hline
    $E_{\mathrm SPH}$&   $3.71\times10^{-1}$        &$1.63\times10^{-1}$       &$8.2\times10^{-2}$         &$3.5\times10^{-2}$        &$1.8\times10^{-2}$       &$1.1\times10^{-2}$     &$8.9\times10^{-3}$\\
    \hline
    $P_{\mathrm CSPM}$&   $5.3\times10^{-2}$        &$1.1\times10^{-2}$       &$3.3\times10^{-3}$        &$5.2\times10^{-4}$        &$1.3\times10^{-4}$       &$5.1\times10^{-5}$  &$3.2\times10^{-5}$\\
    \hline
    $E_{\mathrm CSPM}$&   $1.6\times10^{-2}$        &$4.4\times10^{-3}$       &$1.5\times10^{-3}$        &$2.4\times10^{-4}$        &$5.8\times10^{-5}$        &$5.9\times10^{-5}$   &$3.7\times10^{-5}$ \\
    \hline
    $P_{\mathrm FPM}$&   $9.6\times10^{-2}$        &$9.3\times10^{-3}$       &$6.5\times10^{-4}$        &$6.4\times10^{-5}$         &$1.3\times10^{-5}$        &$4.8\times10^{-6}$  &$3.0\times10^{-6}$\\
    \hline
    $E_{\mathrm FPM}$&   $6.6\times10^{-2}$        &$8.0\times10^{-3}$       &$6.1\times10^{-4}$        &$5.7\times10^{-5}$         &$9.0\times10^{-6}$        &$3.9\times10^{-6}$  &$2.1\times10^{-6}$\\
    \hline
    $P_{\mathrm SFPM}$&  $5.1\times10^{-2}$        &$1.1\times10^{-2}$       &$1.4\times10^{-3}$        &$1.9\times10^{-4}$        &$4.5\times10^{-5}$        &$1.6\times10^{-5}$  &$9.7\times10^{-6}$\\
    \hline
    $E_{\mathrm SFPM}$&   $4.2\times10^{-2}$       &$1.0\times10^{-2}$       &$1.4\times10^{-3}$        &$1.6\times10^{-4}$        &$3.3\times10^{-5}$        &$1.2\times10^{-5}$  &$6.4\times10^{-6}$\\ 
    \hline
    $P_{\mathrm FDM}$&   $3.4\times10^{-2}$        &$8.4\times10^{-3}$       &$2.1\times10^{-3}$        &$3.3\times10^{-4}$        &$8.4\times10^{-5}$       &$3.3\times10^{-5}$   &$2.1\times10^{-5}$\\
    \hline
    $E_{\mathrm FDM}$&   $1.4\times10^{-2}$        &$3.6\times10^{-3}$       &$1.3\times10^{-3}$        &$1.7\times10^{-4}$        &$2.7\times10^{-5}$       &$8.0\times10^{-6}$    &$3.5\times10^{-6}$\\
    \hline
\end{tabular}
\end{table}

It is observed in Tab.\ref{AreEffTable}, for all the methods, the calculated $P$ monotonically decrease with increasing SPH particles.
This is as expected, which indicates that the precision improves as the particle number increases. 
Also, the efficiency, which is defined as an overall evaluation of speed and accuracy, improves as the SPH particle increase.
The modified SPH methods included in this study; namely, CSPM, FPM, and SFPM show significantly better performance than that of the standard SPH method.
It is found that as the number of SPH particle further increases, the improvement of efficiency gradually slows down. 
This happens due to that the precision of the interpolation starts to be saturated while the computational time increases linearly.
It is further expected for the higher dimensional case since the computation cost increases much faster with increasing particle number, the efficiency of the methods may even start to decrease as the particle number becomes significantly large.

\section{IV. Burgers' equation with dynamical SPH particles: on the stability of the algorithm} \label{Sect4}

\begin{figure}[ht]
\begin{tabular}{cc}
\vspace{-30pt}
\begin{minipage}{225pt}
\centerline{\includegraphics[width=300pt]{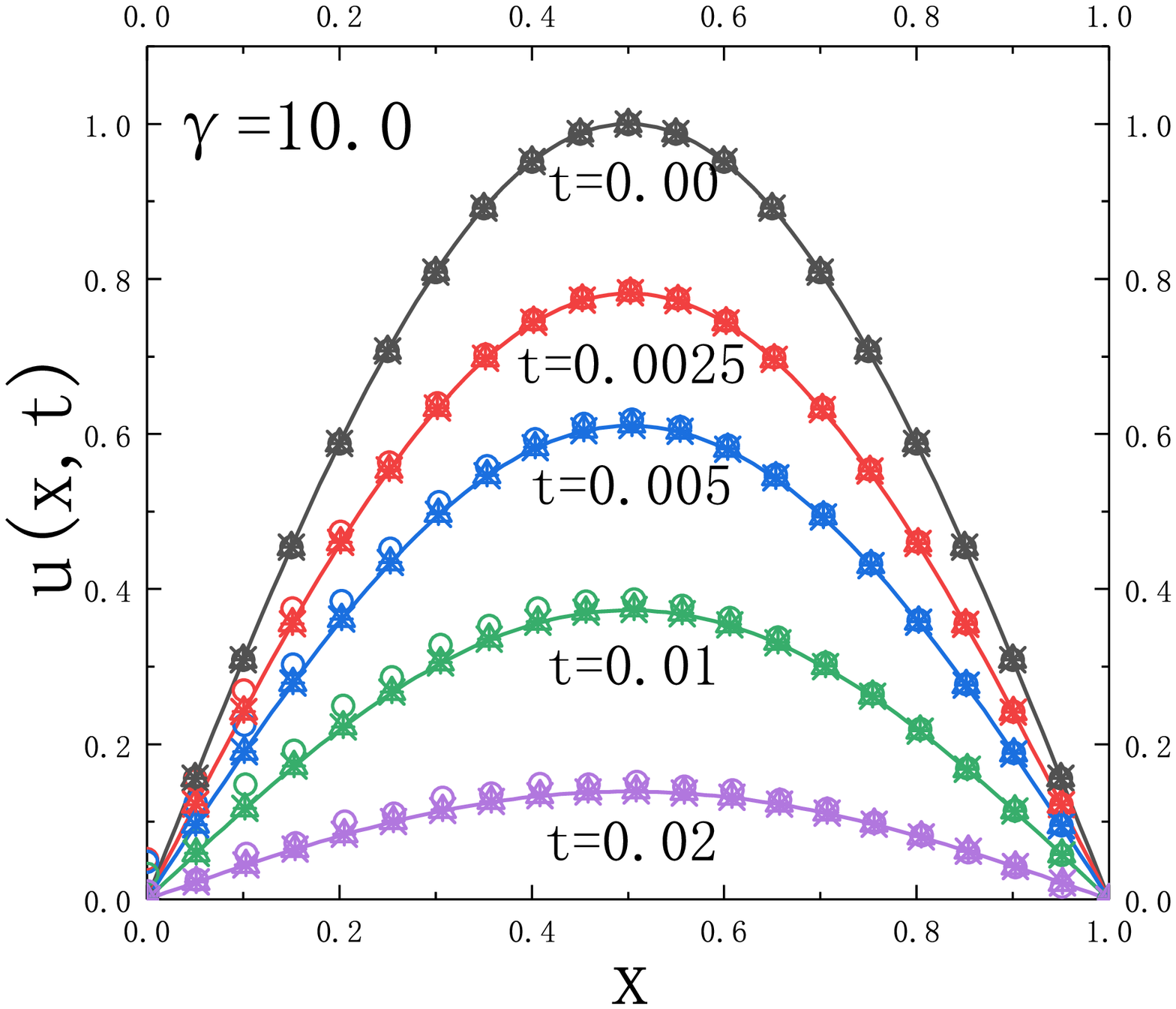}}
\end{minipage}
&
\begin{minipage}{225pt}
\centerline{\includegraphics[width=300pt]{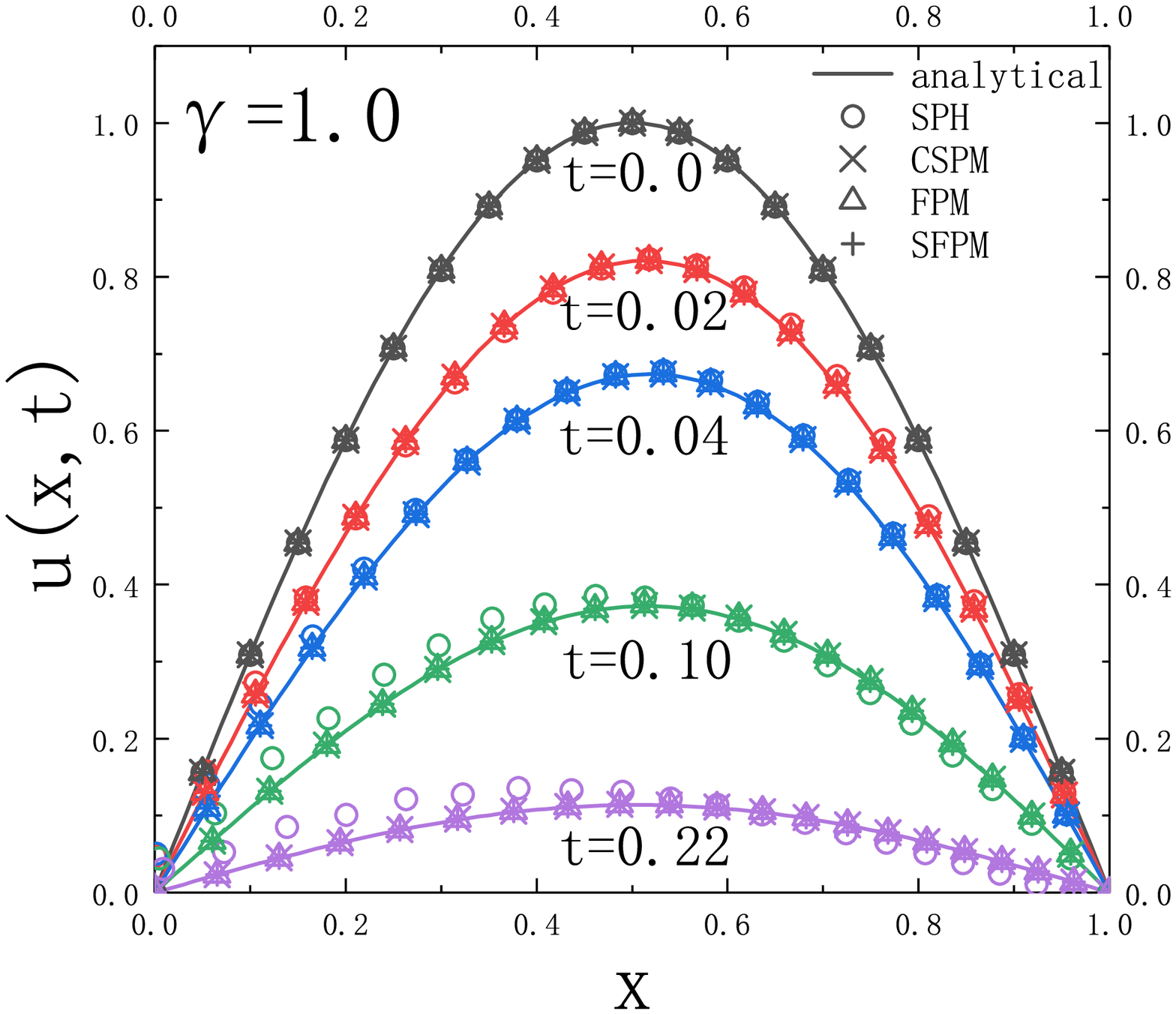}}
\end{minipage}
\\
\begin{minipage}{225pt}
\centerline{\includegraphics[width=300pt]{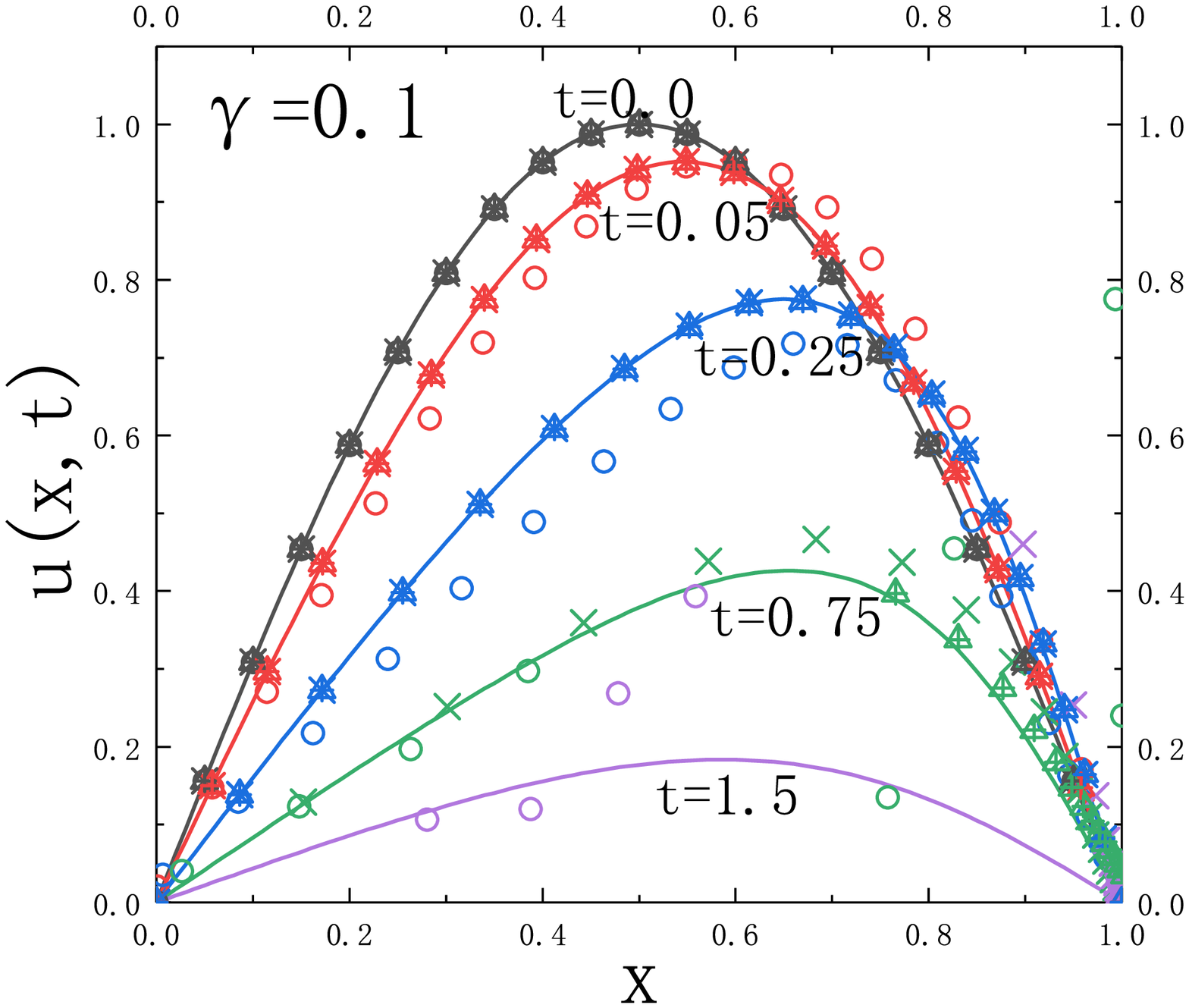}}
\end{minipage}
&
\begin{minipage}{225pt}
\centerline{\includegraphics[width=300pt]{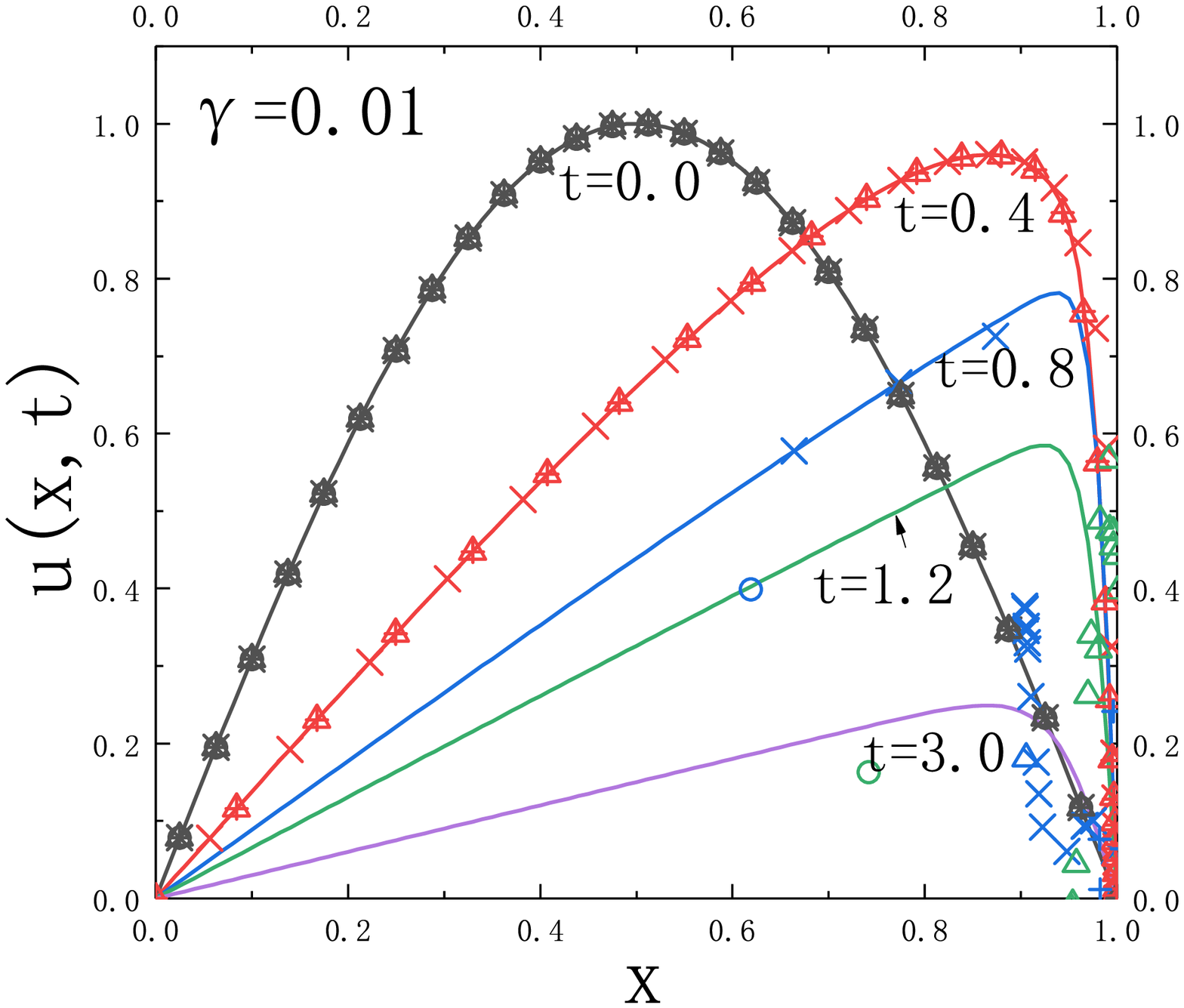}}
\end{minipage}
\end{tabular}
\renewcommand{\figurename}{Fig.}
\caption{(Color online) The numerical solutions of the Burgers' equation in comparison with the analytic ones. 
The calculations are carried out by using different values of the viscosity coefficients for standard SPH, CSPM, FPM, FDM, and SFPM, respectively.
The boundary condition is implemented by using fixed SPH particles.}
\label{temporalEvo-2}
\end{figure}

\begin{figure}[ht]
\begin{tabular}{cc}
\vspace{-30pt}
\begin{minipage}{225pt}
\centerline{\includegraphics[width=300pt]{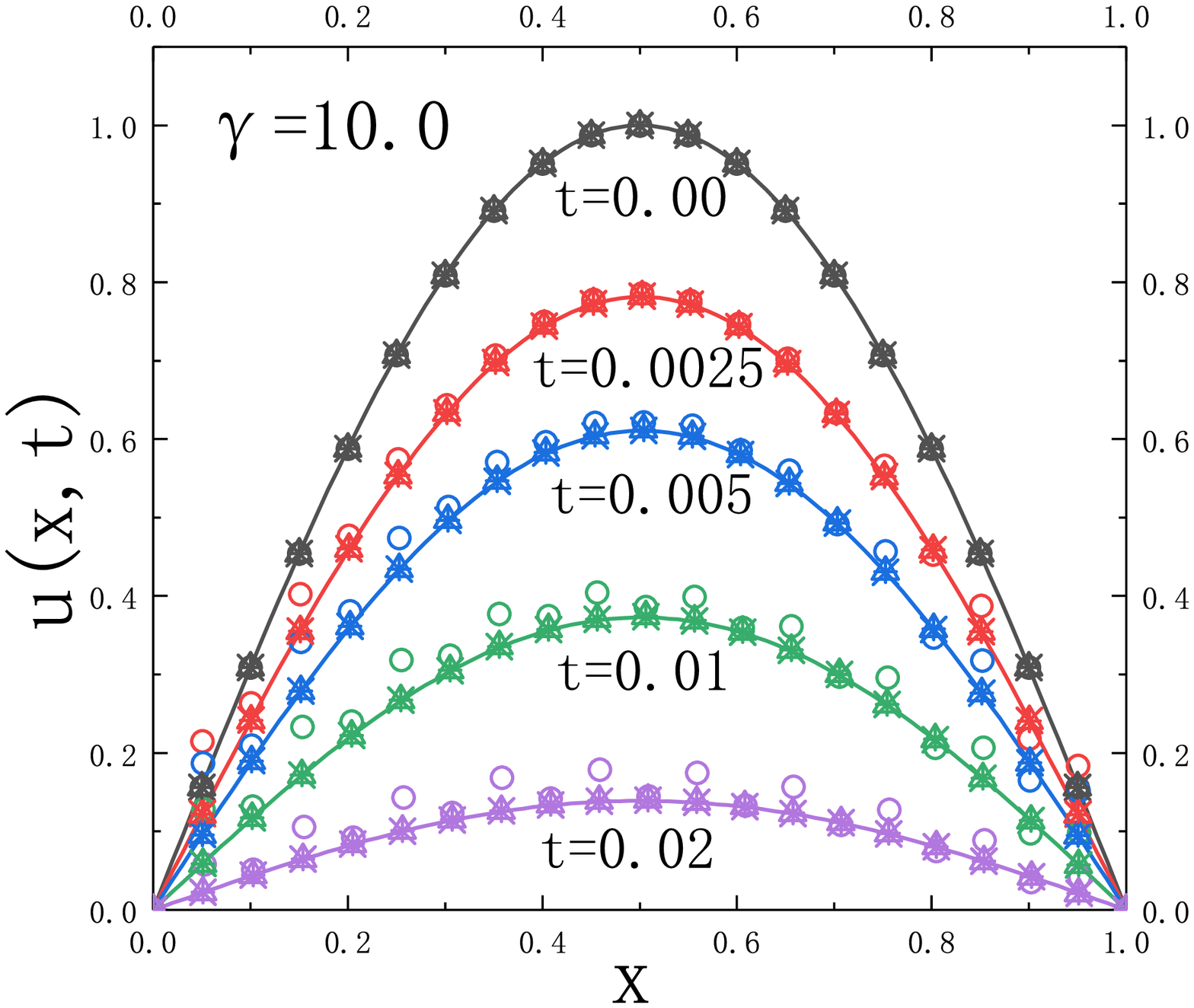}}
\end{minipage}
&
\begin{minipage}{225pt}
\centerline{\includegraphics[width=300pt]{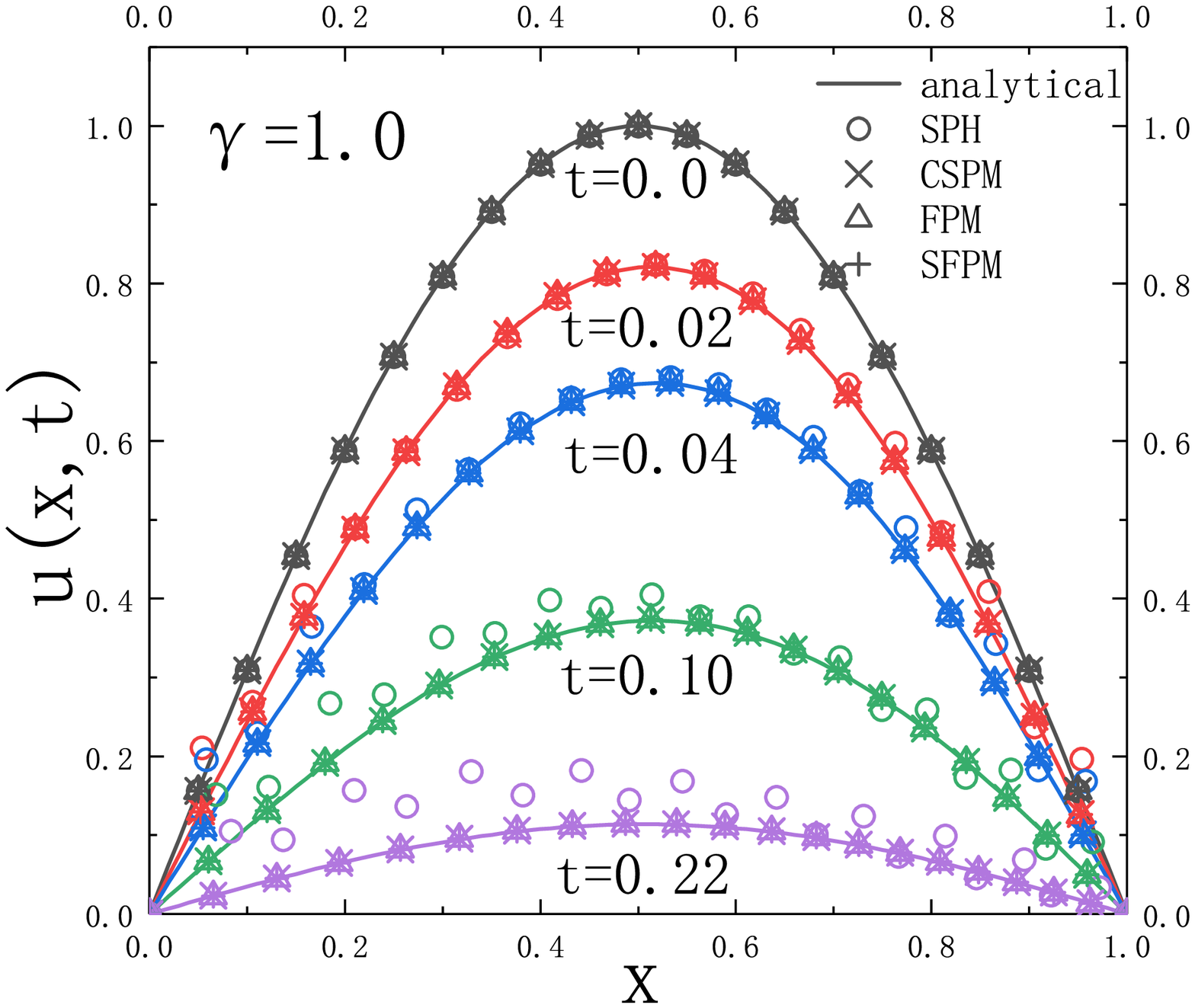}}
\end{minipage}
\\
\begin{minipage}{225pt}
\centerline{\includegraphics[width=300pt]{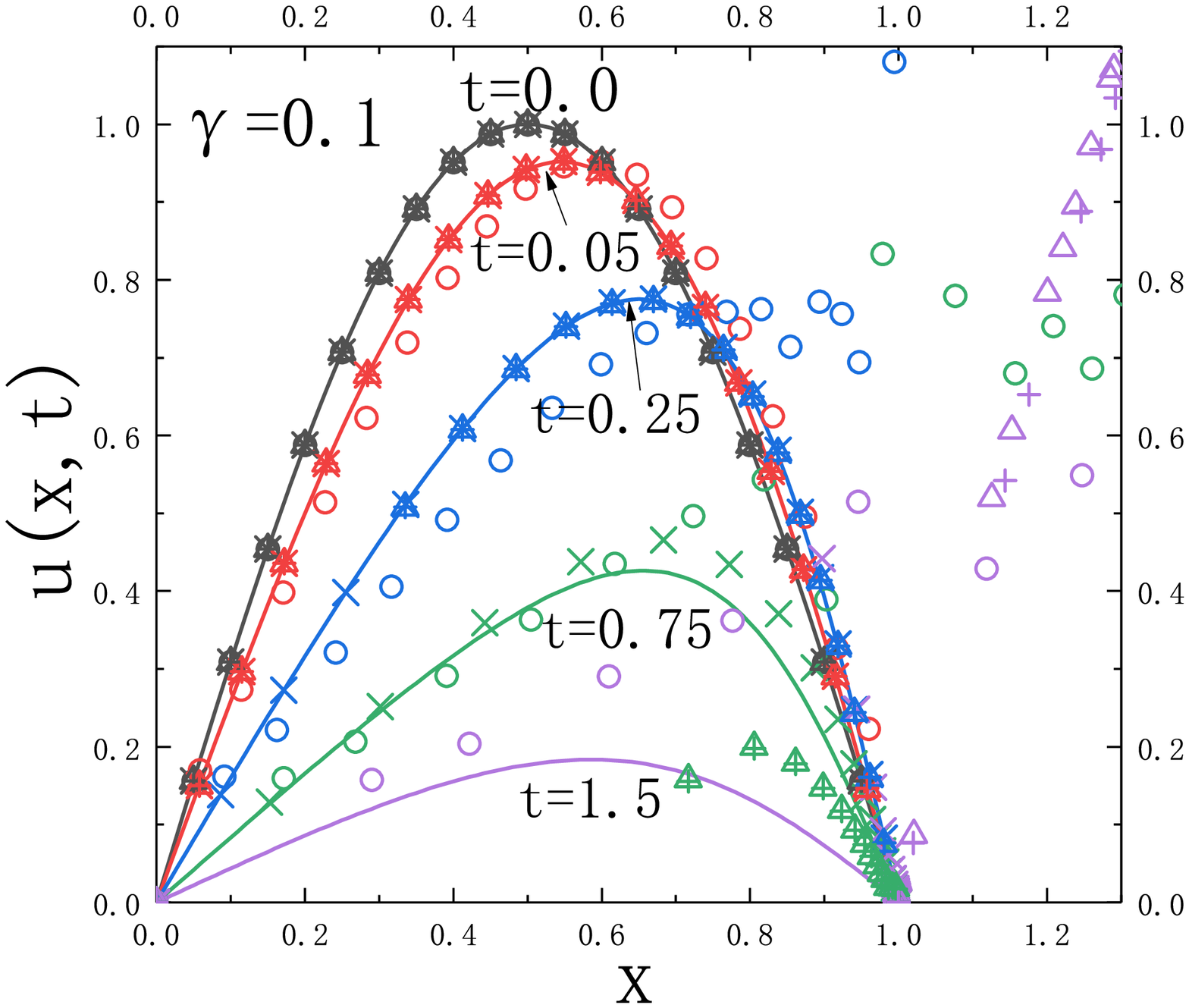}}
\end{minipage}
&
\begin{minipage}{225pt}
\centerline{\includegraphics[width=300pt]{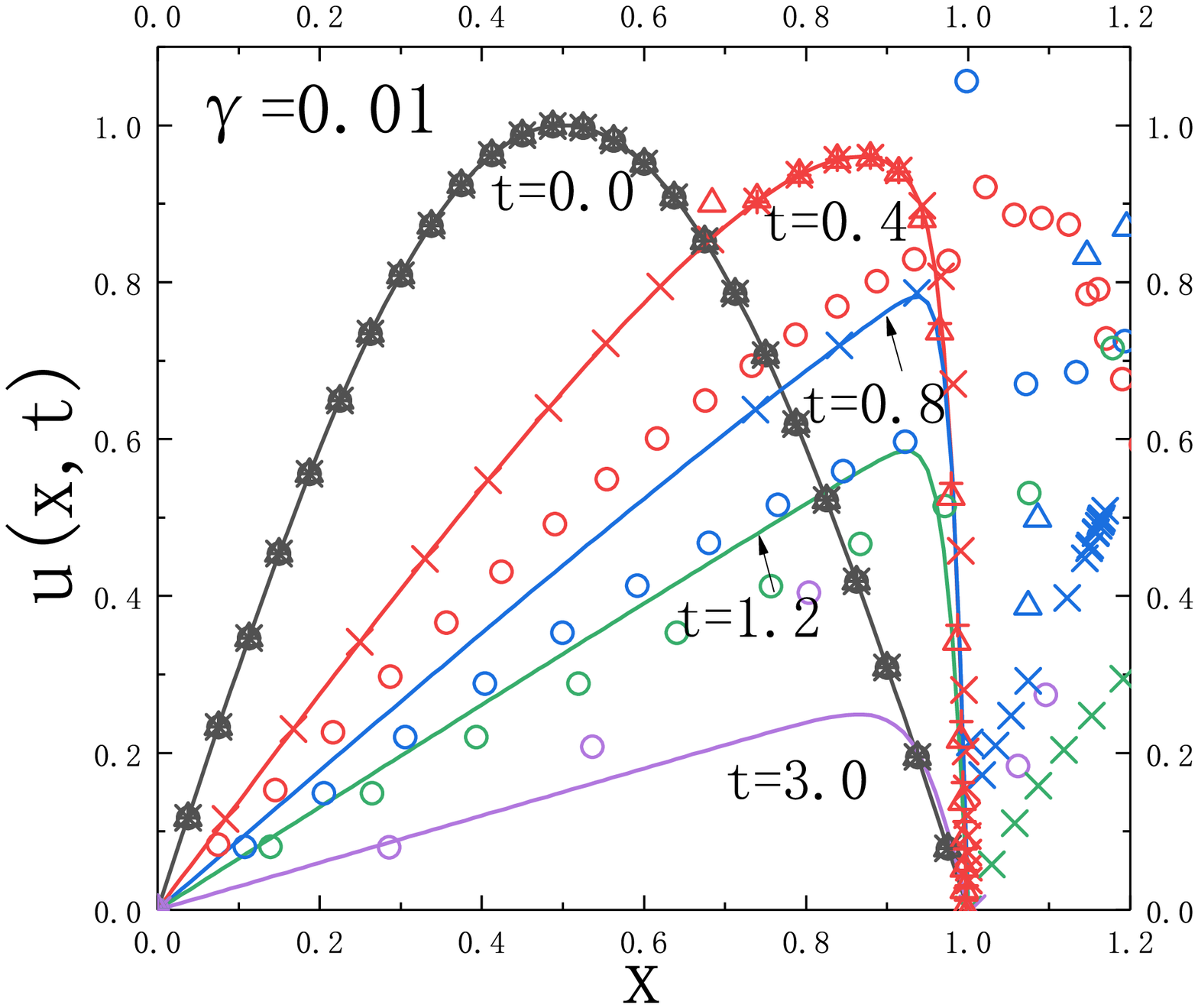}}
\end{minipage}
\end{tabular}
\renewcommand{\figurename}{Fig.}
\caption{(Color online) The numerical solutions of the Burgers' equation in comparison with the analytic ones. 
The calculations are carried out by using different values of the viscosity coefficients for standard SPH, CSPM, FPM, FDM, and SFPM, respectively.
The boundary condition is implemented by adding extra SPH particles beyond the physical boundary.}
\label{temporalEvo-3}
\end{figure}

\begin{figure}[ht]
\begin{tabular}{cc}
\vspace{-30pt}
\begin{minipage}{225pt}
\centerline{\includegraphics[width=300pt]{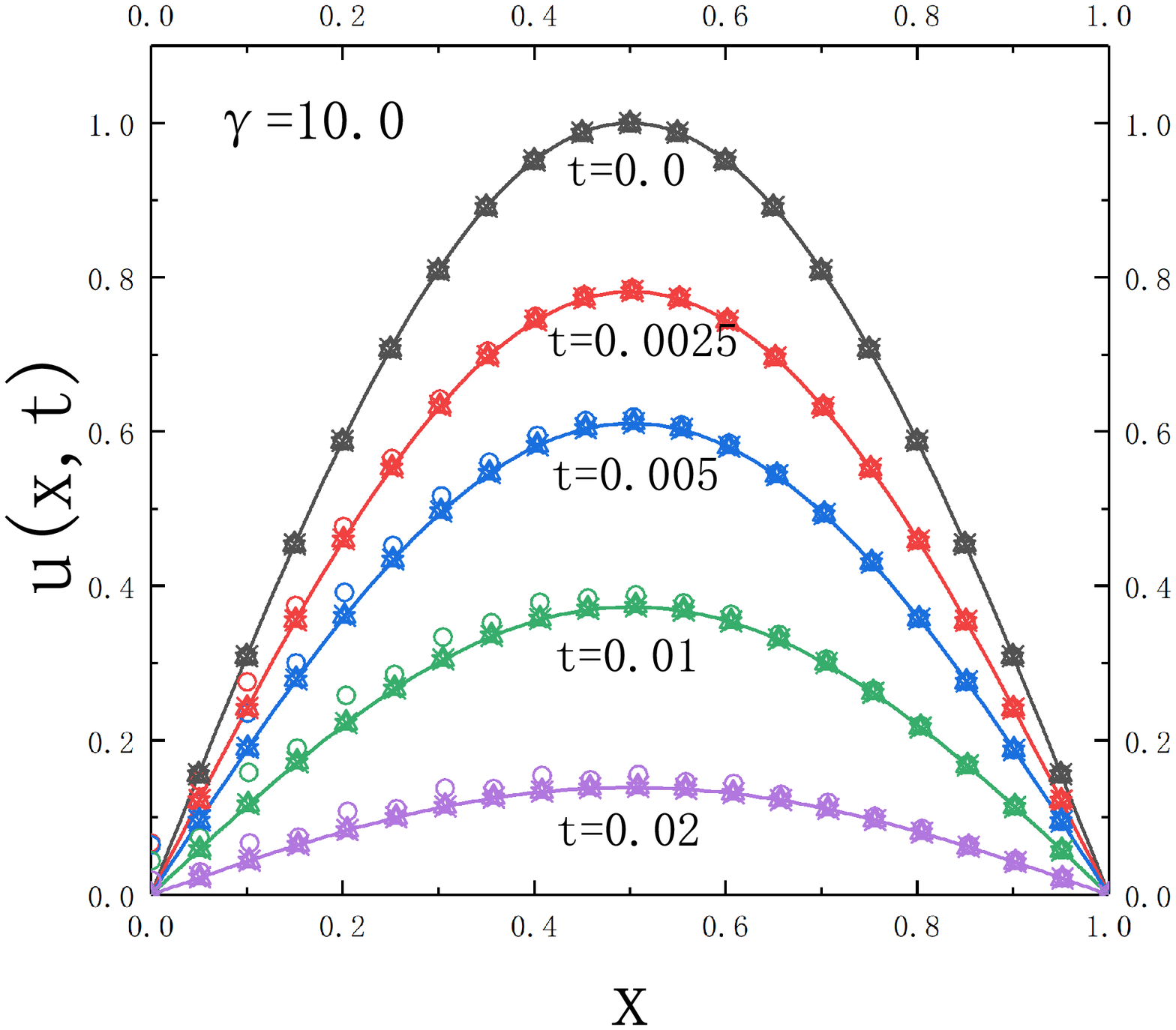}}
\end{minipage}
&
\begin{minipage}{225pt}
\centerline{\includegraphics[width=300pt]{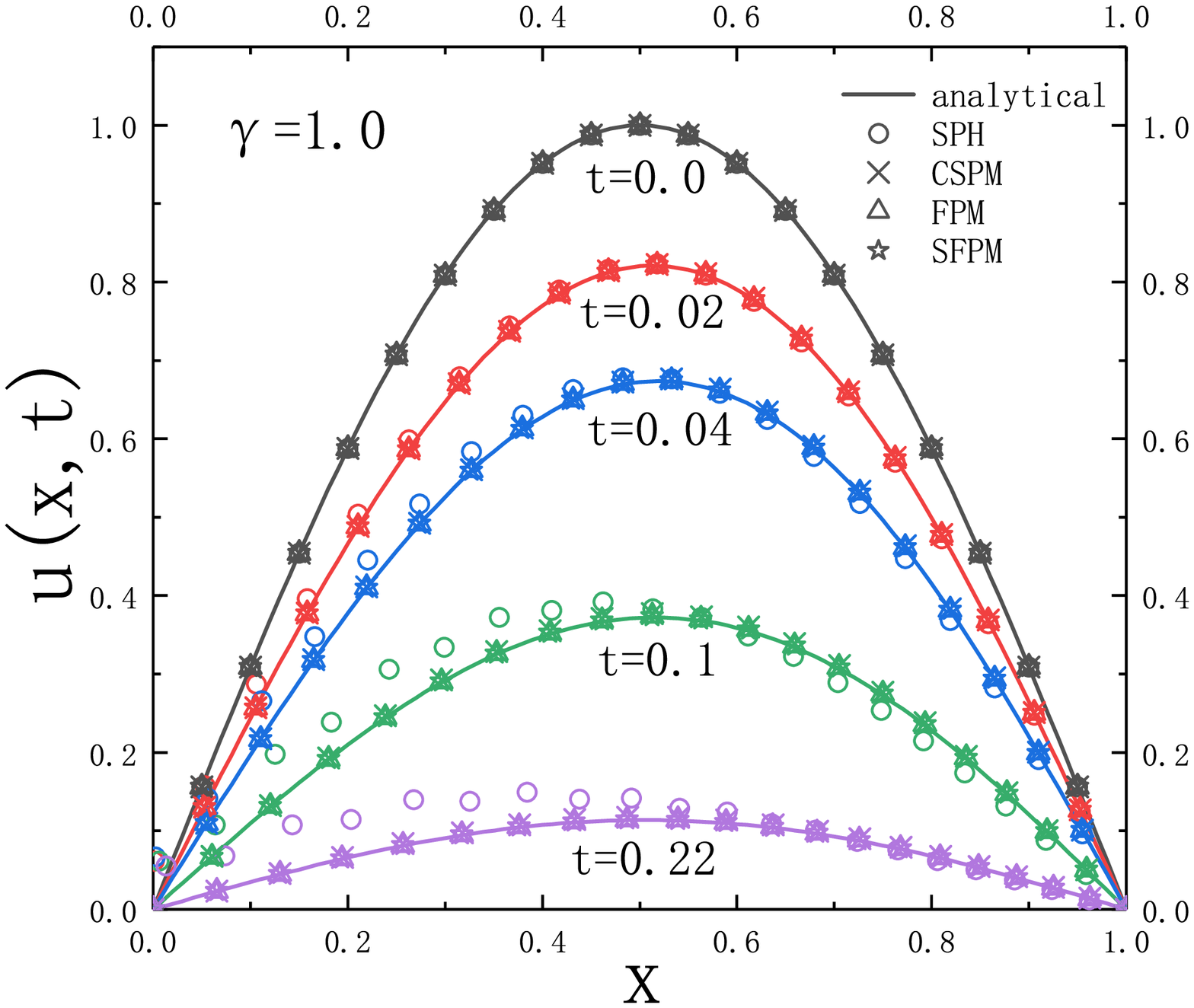}}
\end{minipage}
\\
\begin{minipage}{225pt}
\centerline{\includegraphics[width=300pt]{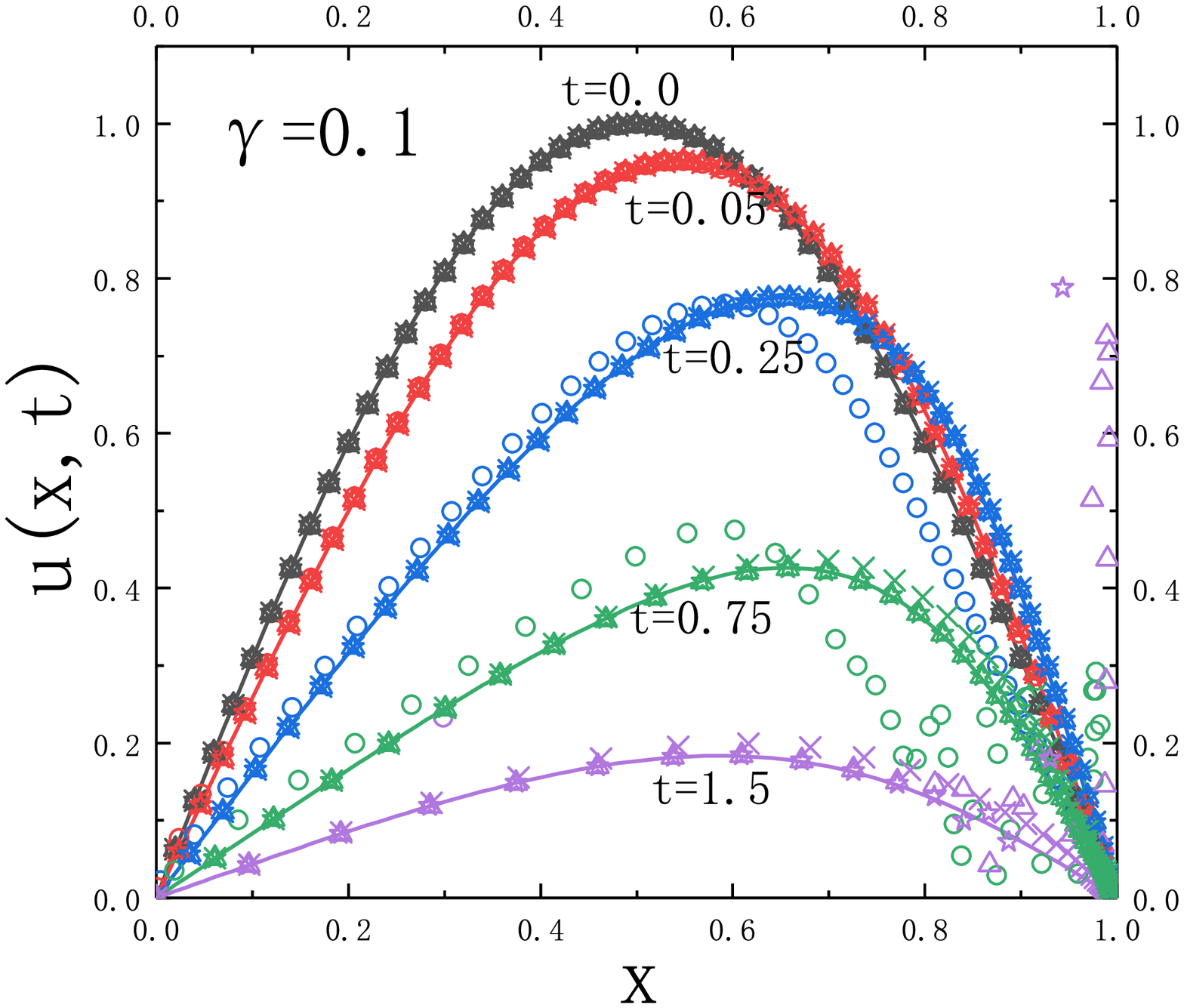}}
\end{minipage}
&
\begin{minipage}{225pt}
\centerline{\includegraphics[width=300pt]{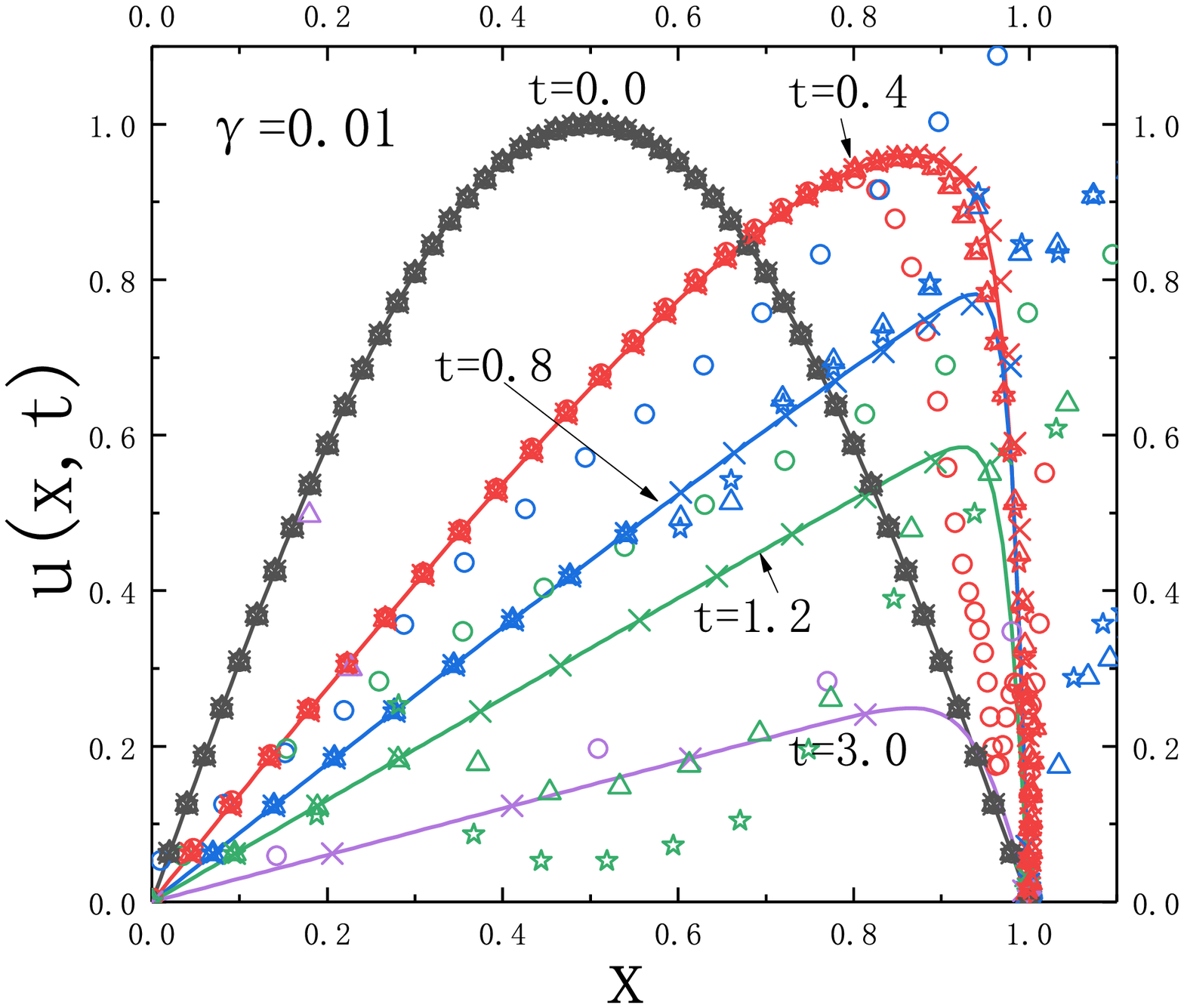}}
\end{minipage}
\end{tabular}
\renewcommand{\figurename}{Fig.}
\caption{((Color online) The numerical solutions of the Burgers' equation in comparison with the analytic ones. 
The calculations are carried out by using different values of the viscosity coefficients for standard SPH, CSPM, FPM, FDM, and SFPM, respectively.
The boundary condition is implemented by adding mirrored imaginary SPH particles beyond the boundary.}
\label{temporalEvo-4}
\end{figure}

\begin{figure}[ht]
\begin{tabular}{cc}
\vspace{-26pt}
\begin{minipage}{225pt}
\centerline{\includegraphics[width=225pt]{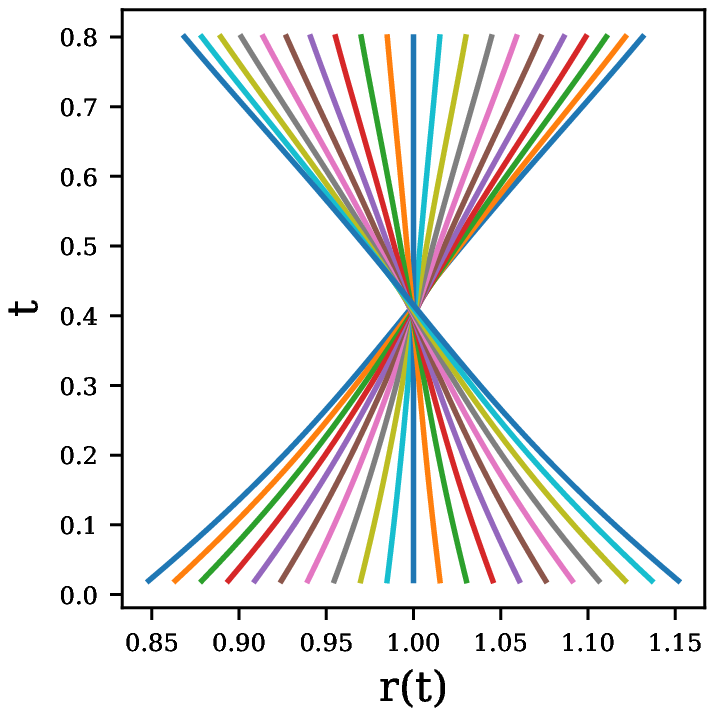}}
\end{minipage}
&
\begin{minipage}{225pt}
\centerline{\includegraphics[width=225pt]{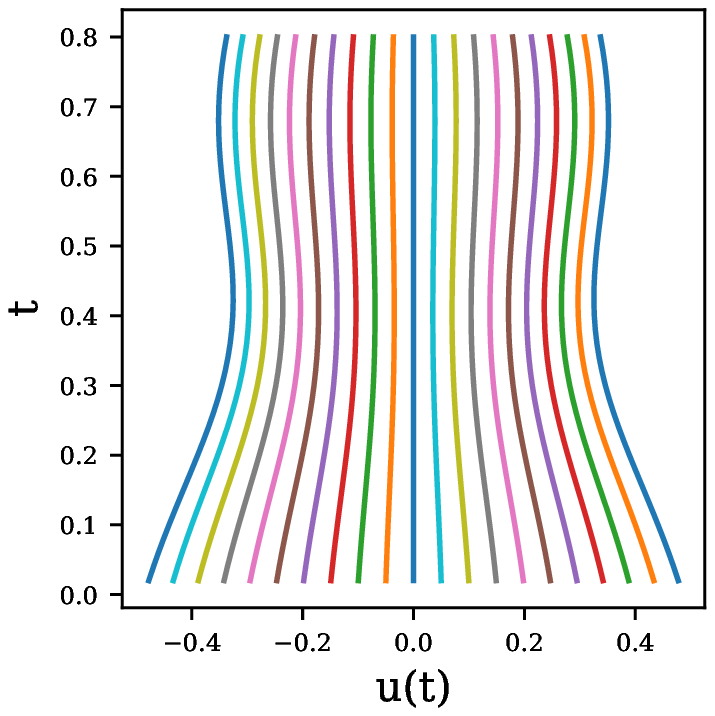}}
\end{minipage}
\\
\vspace{-26pt}
\begin{minipage}{225pt}
\centerline{\includegraphics[width=225pt]{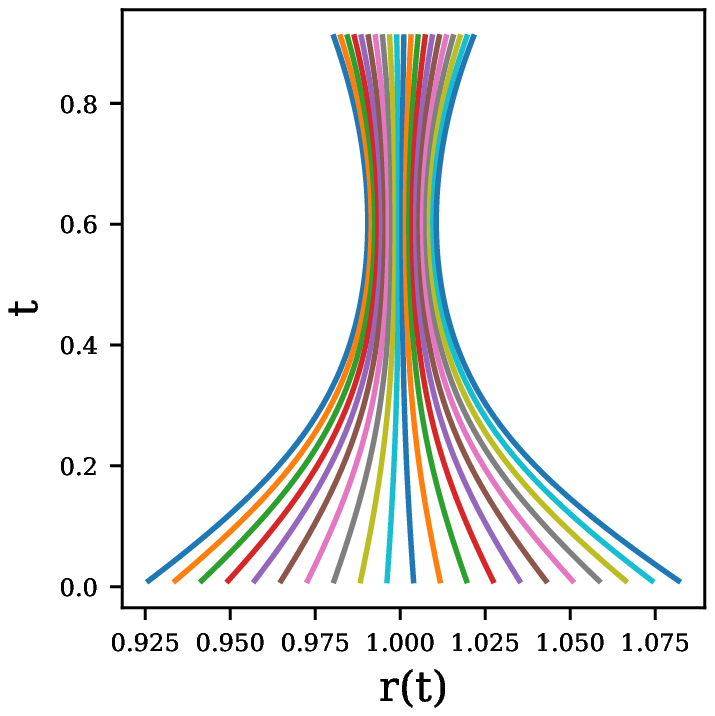}}
\end{minipage}
&
\begin{minipage}{225pt}
\centerline{\includegraphics[width=225pt]{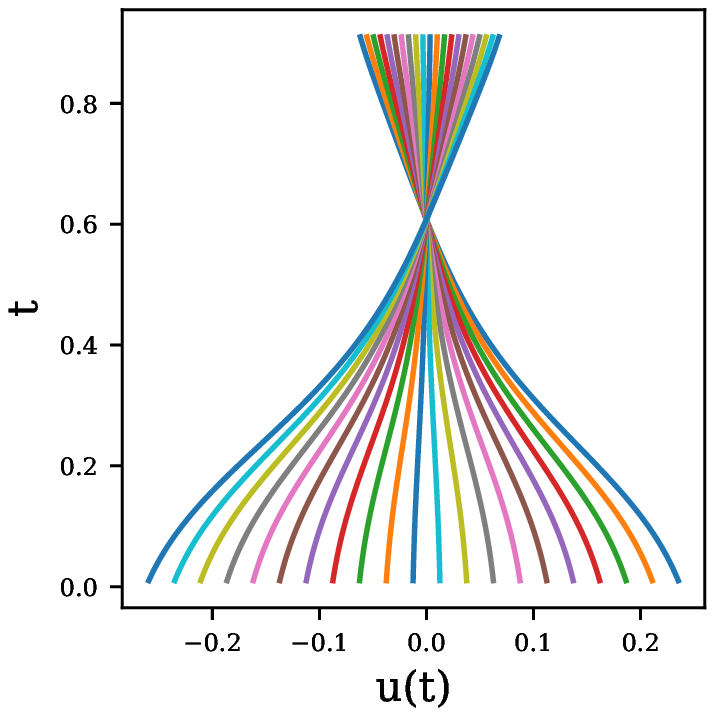}}
\end{minipage}
\\
\vspace{-17pt}
\begin{minipage}{225pt}
\centerline{\includegraphics[width=225pt]{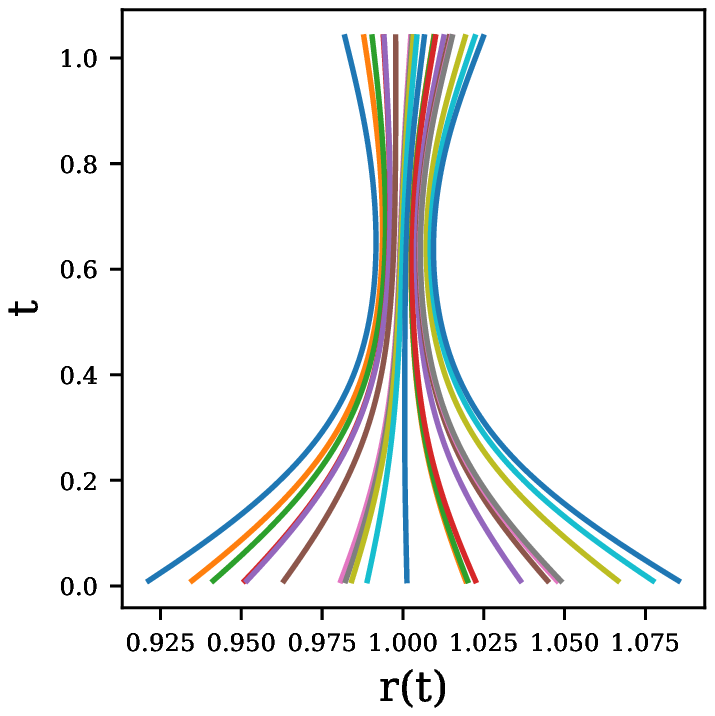}}
\end{minipage}
&
\begin{minipage}{225pt}
\centerline{\includegraphics[width=225pt]{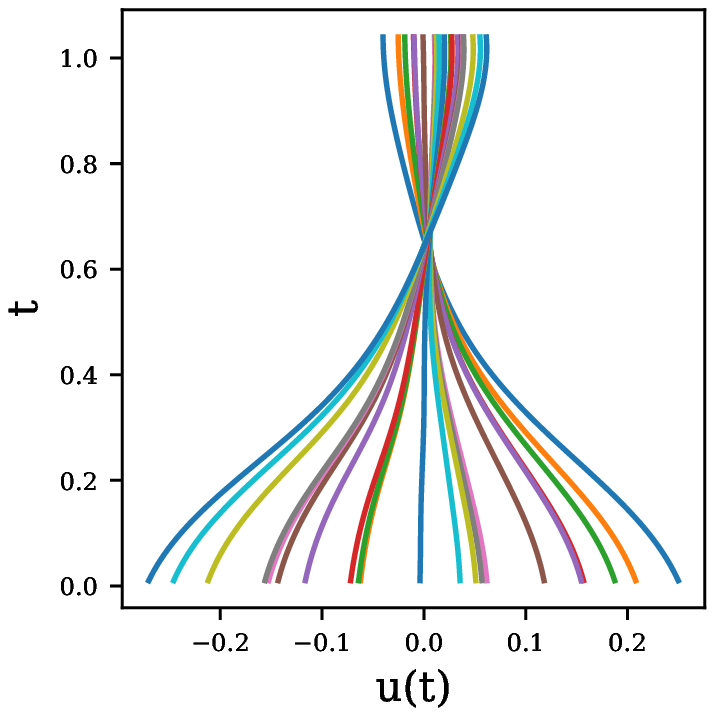}}
\end{minipage}
\end{tabular}
\renewcommand{\figurename}{Fig.}
\caption{(Color online) The numerical solutions of the Burgers' equation for standard SPH by employing the boundary condition with mirrored imaginary SPH particles.
The plots show the trajectories of the positions (on the l.h.s.) and velocities (on the r.h.s.) of SPH particles, by curves in different colors.
Here the calculations are carried out for the range $x\in (0,2)$, while the results are only shown in the vicinity region of $x\sim 1$.
}
\label{fig-particle-traverse}
\end{figure}

This section is devoted to the stability analysis of different schemes regarding the boundary condition and initial particle distribution. 
When not appropriately configured, a numerical algorithm might be plagued by instability.
Here we investigate particularly how the boundary condition may affect the numerical instability.
As discussed in the introduction, there are at least three different types of characteristics of the SPH algorithm closely related to stability issue, namely, self-regularization, pairing instability, and tensile instability.
All of them are related to the particle distribution and the specific form of the kernel function.
SPH can automatically maintain a reasonable particle arrangement~\cite{sph-algorithm-review-07,sph-algorithm-review-09} during the course of the temporal evolution.
Owing to the form of the kernel function, mutual repelling forces are exerted between neighboring SPH particles, which automatically drives the system towards a regular particle distribution.
Intuitively, it corresponds to ``re-meshing" procedure that is utilized in other Lagrangian mesh methods.
As the SPH particles come closer, paring instability takes place as soon as the repelling force start to decrease with decreasing neighbor distance.
Furthermore, under certain conditions~\cite{sph-algorithm-tensile-insti-01,sph-algorithm-tensile-insti-02} tensile instability takes place and disrrupts the particles distribution.
In what follows, we attempt to configure an initial SPH distribution where the distance of neighboring SPH particles is randomized but within the range where particles repelling each other and the strength decreases with increasing distance.
However, our calculation does not reveal any automatical re-meshing.
Moreover, we show that boundary condition may introduce small perturbation which is accumulated over time and turned into oscillations.
Such instability is observed to be accompanied by particle crossing.
In order to carry out the analysis, one allows the SPH particle to move according to the continuity equation, namely, Eq.(\ref{rhoEq}). 
In other words, we will solve Eqs.(\ref{BurgerEq}) and (\ref{rhoEq}) simultaneously, as in most SPH implementations.
The numerical results are presented in Figs.\ref{temporalEvo-2}-\ref{fig-particle-traverse}.

A critical aspect of the present study is to show that the boundary condition plays an essential role in the SPH algorithm.
From its appearance, the boundary condition Eq.(\ref{BoundaryCondition}) of the problem looks ``benign". 
However, it turns out that the numerical stability sensitively depends on its implementation.
We study three types of implementation for the boundary condition.
First, we freeze the coordinates of the SPH particles on the boundary.
Secondly, we extend the system beyond the boundary by adding more SPH particles and let those additional particles evolve freely.
To be specific, we do not explicitly exert any additional condition for the most outward particles.
Lastly, we implement mirroring boundary condition, by adding imaginary particles outside of the boundary to ensure the validity of Eq.(\ref{BoundaryCondition}).
To be more precise, the imaginary particles are mirror images; namely, their positions are mirrored with respect to the location of the boundary, and their velocities are inverted regarding their counterparts.
The obtained results are shown in Figs.\ref{temporalEvo-2}-\ref{temporalEvo-4}.

The calculated results indicate that it is essential to appropriately implement the boundary condition, determined by Eq.(\ref{BoundaryCondition}).
As shown in Fig.\ref{temporalEvo-2}, the calculations with fixed boundary SPH particles start to fail as the coefficient of viscosity becomes significantly small.
As $\gamma$ decreases, it takes a longer time for the system to evolve, and subsequently, small deviations are accumulated for an extended period, and the results demonstrate more significant discrepancy.
It is observed that most schemes show significant deviations for $\gamma< 0.1$ and for $t>0.75$, where the standard SPH fails dramatically.
While one adds additional SPH particles beyond the boundary of the physical problem and apply free boundary condition for those particles, the results show a very similar feature as $\gamma$ becomes sufficiently small.
This can be observed in Fig.\ref{temporalEvo-3}.
In Fig.\ref{temporalEvo-4}, the results are shown by employing mirrored boundary condition.
We did not draw the mirrored SPH particles beyond the boundary as they are symmetrically positioned regarding their counterparts.
In general, the results are significantly improved.
It is found that the implemented boundary condition gives a significant impact on the resulting temporal evolution.
For CSPM, the results are more stable and agree reasonably with the analytic ones for all different viscosity coefficients.
Although, in some region, the numerical results are still found to deviate from the analytic ones.
For FPM and SFPM, small oscillations are observed when the time becomes more substantial as SPH particles start to pile up at the right boundary $x=1$.
For the standard SPH method, the results are less satisfactory compared to the other approaches.
We understand that it is mostly related to the fact that the standard SPH method suffers a severe problem with kernel and particle consistency.

In order to investigate the instability occurs near the rightmost boundary, we carry out further calculations focusing the region near the vicinity of $x=1$.
To be specific, the numerical calculations are carried out for the range of $x\in (0,2)$ with the boundary being $x=0$ and $x=2$.
The imaginary SPH particles beyond the original boundary are now treated as real particles and evolve according to the Burgers' equation. 
In this context, the particles on both sides are moving toward the center at $x=1$.
The numerical results are shown in Fig.\ref{fig-particle-traverse}.
The two plots on the first row are the trajectories the positions and velocities of SPH particles, shown by curves in different colors.
A total of 125 SPH particles are initially distributed uniformly, where $h=0.2$ and the size of the time step $dt=0.02$.
The plots on the second as well as third rows are the same as the first row, except for the following difference.
For the second row, the SPH particle distribution is also uniform initially, but a total number of 250 particles, with $h=0.1$ and $dt=0.01$.
For the third row, regarding the initial condition for the case of the second row, the positions and velocity of SPH particles are perturbed by a small amount at the initial time. 
While the calculations are carried out for the range $x\in (0,2)$ with the boundary are defined by the points $x=0$ and $2$, the results are only shown in the vicinity region of $x\sim 1$.
It is found that on the first row of Fig.\ref{fig-particle-traverse}, particles simple travel across each other after they come close.
This is a troublesome result, especially when comparing to the analytic results, the fluid on either side of the boundary should never traverse $x=1$.
Therefore, some fundamental problem regarding interpolation is implied.
As will be discussed below, the stability mentioned above in Figs.\ref{temporalEvo-2}-\ref{temporalEvo-4} is likely related to the present phenomenon.
On the second and third rows, particles bound back from each other due to the repelling force between them, as intended.
As discussed before, if particles maintain at an appropriate distance between each other, particle auto-remeshing is expected.
In the plots on the second row of Fig.\ref{fig-particle-traverse}, this is manifested, to a certain degree, in terms of the fact that all the SPH particles gradually decelerate and come to a full stop simultaneously at $t\sim 0.6$. 
Further, in the plots on the third row, it is shown that the particles also bounce back from each other, owing to the auto-remeshing mechanism.
However, our results indicate that auto-remeshing does not always work in the present context, in particular, where the fluid becomes piled up at $x=1$ and is forced to squeeze into a small region, as governed by the Burgers' equation. 
In this case, the results become quite sensitive to the specific parameter used in the calculations, as demonstrated by the plots on the first and second rows.
When the auto-remeshing fails, as shown in the first row, the SPH particles are found to traverse each other when they stay too close to each other.
Judging from the form of the kernel function, when $\Delta x < h$, the repelling between two neighboring SPH particles decrease with decreasing distance.
Therefore, as the neighboring particles become too close, as it inevitably happens in the case of the Burgers' equation, the mechanism of particle auto-remeshing may fail.
In fact, we observe that the resultant particle traversing occurs for a broad choice of parameters in the present scenario of the Burgers' equation.
We note that such a phenomenon is observed to accompany the instability discussed previously concerning Figs.\ref{temporalEvo-2}-\ref{temporalEvo-4} closely.

\section{V. Concluding remarks} \label{Sect5}

To summarize, in this paper, we carry out an analysis regarding the stability, precision, and efficiency of the SPH method.
In particular, various aspects of the algorithm are explored such as the boundary condition, initial particle distribution, and interpolation scheme.
Besides, a generalized FPM scheme, the symmetrized finite particle method, is employed in our study.
The main advantage of the SFPM is that its implementation does not explicitly involve the derivatives of the kernel function.
We studied two different scenarios for the implementation of the equation of motion, where the SPH particles are either stationary or dynamically evolved in time.
In the first scenario, the obtained results are also compared with those obtained by using the FDM as well as analytic ones.
All numerical schemes provide reasonable results in the case.
Among others, it is indicated that CSPM and SFPM provides an overall better performance regarding precision as well as stability. 

For the second scenario, the positions and velocity of the SPH particles are determined by the equations of motion.
Though it is more meaningful to apply this scheme to topics with dynamical boundary conditions, in the present study, it is adopted to the same problem of the Burgers' equation.
It is subsequently found that the boundary condition implementation plays a crucial role in the success of the algorithm.
Among the three implementations, the mirrored boundary condition leads to mostly satisfactory results.
Still, instability in terms of oscillations is observed when particle crossing starts to take place.
Even for the case of mirrored boundary condition, increasing oscillation is observed for FPM and SFPM methods in regions with more significant particle density.
Since the SFPM does not explicitly involve any derivative of the kernel function, the observed instability is likely of a distinct nature.
This is because the sufficient condition of tensile instability involves the second order of the kernel function when derived regarding small perturbations~\cite{sph-algorithm-tensile-insti-01}.
One crucial feature of observed instability is that the oscillations take place when SPH particles start to cross each other.
We will try to address the issue of particle crossing and the related instability in future work.

\section*{Acknowledgements}

We gratefully acknowledge the financial support from Brazilian funding agencies Funda\c{c}\~ao de Amparo \`a Pesquisa do Estado de S\~ao Paulo (FAPESP), 
Conselho Nacional de Desenvolvimento Cient\'{\i}fico e Tecnol\'ogico (CNPq), Coordena\c{c}\~ao de Aperfei\c{c}oamento de Pessoal de N\'ivel Superior (CAPES), 
and National Natural Science Foundation of China (NNSFC) under contract No. 11805166.

\bibliographystyle{h-physrev}
\bibliography{references_qian}

\begin{thebibliography}{10}

\bibitem{sph-app-astro-01}
L.~Lucy,
\newblock Astrophys. J. {\bf 82} (1977) 1013.

\bibitem{sph-app-astro-02}
R.~Gingold and J.~Monaghan,
\newblock Mon. Not. R. Astro. Soc. {\bf 181} (1977) 375.

\bibitem{sph-algorithm-review-01}
J.~J. {Monaghan},
\newblock Ann. Rev. Astron. Astrophys. {\bf 30} (1992) 543.

\bibitem{sph-algorithm-review-02}
W.~Benz,
\newblock {\em {Smooth Particle Hydrodynamics - a Review }}NATO SAI Series C:
  Mathematical and Physical Sciences (Kluwer Academic Publishers, 1990).

\bibitem{sph-algorithm-review-03}
J.~J. {Monaghan},
\newblock Rep. Prog. Phys. {\bf 68} (2005) 1703.

\bibitem{sph-algorithm-review-04}
J.~J. {Monaghan} and D.~J. {Price},
\newblock Mon. Not.Roy. Astr.Soc. {\bf 328} (2001) 381.

\bibitem{sph-algorithm-review-05}
S.~Rosswog,
\newblock New Astron. Rev. {\bf 53} (2009) 78, arXiv:0903.5075.

\bibitem{sph-algorithm-review-06}
M.~B. Liu and G.~R. Liu,
\newblock Archives of Computational Methods in Engineering {\bf 17} (2010) 25.

\bibitem{sph-algorithm-review-10}
V.~Springel,
\newblock Ann. Rev. Astron. Astrophys. {\bf 48} (2010) 391, arXiv:1109.2219.

\bibitem{sph-algorithm-review-09}
S.~Rosswog,
\newblock (2014), arXiv:1406.4224.

\bibitem{sph-app-astro-03}
C.~L. Fryer, G.~Rockefeller, and M.~S. Warren,
\newblock Astrophys. J. {\bf 643} (2006) 292, arXiv:astro-ph/0512532.

\bibitem{sph-app-01}
L.~D. {Libersky} and A.~G. {Petschek},
\newblock  {\bf 395} (1991) 248.

\bibitem{sph-app-02}
J.~Bonet and S.~Kulasegaram,
\newblock Int. J. Numer. Methods Eng. {\bf 47} (2000) 1189.

\bibitem{sph-app-03}
T.~Rabczuk and J.~Eibl,
\newblock Int. J. Numer. Methods Eng. {\bf 56} (2003) 1421.

\bibitem{sph-app-04}
M.~Herreros and M.~Mabssout,
\newblock Comput. Methods Appl. Mech. Engrg. {\bf 200} (2011) 1833.

\bibitem{hydro-review-07}
P.~Romatschke,
\newblock Int. J. Mod. Phys. {\bf E19} (2010) 1, arXiv:0902.3663.

\bibitem{sph-review-1}
Y.~Hama, T.~Kodama, and O.~Socolowski~Jr.,
\newblock Braz.J.Phys. {\bf 35} (2005) 24, arXiv:hep-ph/0407264.

\bibitem{hydro-review-06}
R.~Derradi~de Souza, T.~Koide, and T.~Kodama,
\newblock Prog. Part. Nucl. Phys. {\bf 86} (2016) 35, arXiv:1506.03863.

\bibitem{adscft-fluidgravity-01}
R.~Baier, P.~Romatschke, D.~T. Son, A.~O. Starinets, and M.~A. Stephanov,
\newblock JHEP {\bf 04} (2008) 100, arXiv:0712.2451.

\bibitem{adscft-fluidgravity-02}
R.~Loganayagam,
\newblock JHEP {\bf 05} (2008) 087, arXiv:0801.3701.

\bibitem{adscft-fluidgravity-04}
O.~Aharony, S.~Minwalla, and T.~Wiseman,
\newblock Class. Quant. Grav. {\bf 23} (2006) 2171, arXiv:hep-th/0507219.

\bibitem{adscft-fluidgravity-06}
V.~Lysov and A.~Strominger,
\newblock (2011), arXiv:1104.5502.

\bibitem{adscft-01}
J.~M. Maldacena,
\newblock Int. J. Theor. Phys. {\bf 38} (1999) 1113, arXiv:hep-th/9711200,
\newblock [Adv. Theor. Math. Phys.2,231(1998)].

\bibitem{adscft-02}
E.~Witten,
\newblock Adv. Theor. Math. Phys. {\bf 2} (1998) 253, arXiv:hep-th/9802150.

\bibitem{sph-corr-1}
J.~Takahashi {\em et~al.},
\newblock Phys.Rev.Lett. {\bf 103} (2009) 242301, arXiv:0902.4870.

\bibitem{hydro-vn-2}
D.~Teaney and L.~Yan,
\newblock Phys.Rev. {\bf C86} (2012) 044908, arXiv:1206.1905.

\bibitem{sph-corr-4}
R.~P.~G. Andrade, F.~Grassi, Y.~Hama, and W.-L. Qian,
\newblock Phys.Lett. {\bf B712} (2012) 226, arXiv:1008.4612.

\bibitem{hydro-corr-2}
Z.~Qiu and U.~Heinz,
\newblock Phys.Lett. {\bf B717} (2012) 261, arXiv:1208.1200.

\bibitem{sph-corr-ev-4}
W.-L. Qian, R.~Andrade, F.~Gardim, F.~Grassi, and Y.~Hama,
\newblock Phys.Rev. {\bf C87} (2013) 014904, arXiv:1207.6415.

\bibitem{sph-algorithm-review-07}
D.~J. Price,
\newblock ASP Conf. Ser. {\bf 453} (2012) 249, arXiv:1111.1259.

\bibitem{sph-algorithm-tensile-insti-01}
J.~W. Swegle, D.~Hicks, and S.~Attaway,
\newblock J. Comput. Phys. {\bf 116} (1995) 123.

\bibitem{sph-algorithm-tensile-insti-02}
C.~Dyka and R.~Ingel,
\newblock Comput. \& Struct. {\bf 57} (1995) 573.

\bibitem{sph-algorithm-CSPM-03}
J.~K. Chen, J.~E. Beraun, and C.~J. Jih,
\newblock Computational Mechanics {\bf 23} (1999) 279.

\bibitem{sph-algorithm-pairing-inst-01}
W.~Dehnen and H.~Aly,
\newblock Mon. Not. Roy. Astron. Soc. {\bf 425} (2012) 1068, arXiv:1204.2471.

\bibitem{sph-algorithm-10}
P.~Randles and L.~Libersky,
\newblock Comput. Meth. Appl. M. {\bf 139} (1996) 375.

\bibitem{sph-algorithm-CSPM-01}
J.~Chen and J.~Beraun,
\newblock Comput. Meth. Appl. M. {\bf 190} (2000) 225.

\bibitem{sph-algorithm-FPM-01}
M.~B. Liu and G.~R. Liu,
\newblock Appl. Numer. Math. {\bf 56} (2006) 19.

\bibitem{sph-algorithm-09}
P.~Mota, W.~Chen, and W.-L. Qian,
\newblock Commun. Theor. Phys. {\bf 68} (2017) 382, arXiv:1704.06165.

\bibitem{sph-algorithm-art-viscosity-01}
L.~D. Libersky, A.~G. Petschek, T.~C. Carney, J.~R. HippFirooz, and
  A.~Allahdadi,
\newblock J. Comput. Phys. {\bf 109} (1993) 67.

\bibitem{sph-algorithm-tensile-insti-03}
D.~J. Price,
\newblock J. Comput. Phys. {\bf 231} (2012) 759, arXiv:1012.1885.

\bibitem{sph-algorithm-tensile-insti-04}
T.~Belytschko and S.~Xiao,
\newblock Comput. \& Math. Appl. {\bf 43} (2002) 329.

\bibitem{sph-algorithm-CSPM-02}
J.~K. Chen, J.~E. Beraun, and T.~C. Carney,
\newblock Int. J. Num. Meth. Eng. {\bf 46} (1999) 231.

\bibitem{sph-algorithm-art-viscosity-02}
S.-I. Inutsuka,
\newblock J. Comput. Phys. {\bf 179} (2002) 238.

\bibitem{sph-algorithm-art-viscosity-03}
L.~Cullen and W.~Dehnen,
\newblock MNRAS {\bf 1126} (2010).

\bibitem{sph-algorithm-FPM-05}
C.~Huang, J.~Lei, M.~Liu, and X.~Peng,
\newblock International Journal for Numerical Methods in Fluids {\bf 78} (2015)
  691.

\bibitem{sph-algorithm-08}
J.~P. Morris, P.~J. Fox, and Y.~Zhu,
\newblock J. Comput. Phys. {\bf 136} (1997) 214.

\end{thebibliography}

\end{document}